\documentclass[a4paper,12pt]{article}
\pdfoutput=1
\usepackage[utf8]{inputenc}
\usepackage[T1]{fontenc}

\usepackage{amsmath}
\usepackage{amssymb}
\usepackage{physics}
\usepackage[font=small,labelfont=bf,labelsep=endash,margin=20pt]{caption}
\usepackage[title,header]{appendix}
\usepackage{newtxtext}\usepackage[utopia,greekuppercase=italicized]{mathdesign}\edef\partial{\mathchar\number\partial\noexpand\!}
\usepackage{isomath} 
\newcommand{\updelta}{\deltaup}
\usepackage[margin=25mm]{geometry}
\usepackage{enumitem} 
\setlist{itemsep=2pt,parsep=2pt,topsep=0pt,partopsep=0pt}
\usepackage{graphicx}
\usepackage{url}
\usepackage[hyperref,x11names]{xcolor}
\usepackage[colorlinks=true,linkcolor=Firebrick4,citecolor=blue,urlcolor=SteelBlue4]{hyperref}

\newcommand{\keywords}[1]{\vspace{2mm}\noindent\textbf{Key words:} #1} 
\let\paragraphold\paragraph
\renewcommand*{\paragraph}[1]{\paragraphold{#1.}} 

\newcommand{\upmu}{\muup}
\newcommand{\um}{\ensuremath{\upmu\text{m}}}
\newcommand{\s}{{\text{s}}}

\newcommand{\der}{\mathrm{d}}

\newcommand{\pd}[2]{\frac{\partial #1}{\partial #2}}

\newcommand{\td}[2]{\frac{\der #1}{\der #2}}

\newcommand{\Order}{\mathcal{O}}

\newcommand{\e}{\mathrm{e}}
\newcommand{\avg}[1]{\left\langle #1 \right\rangle}

\begin{document}

\title{\bf Random walk models for the propagation of signalling molecules in one-dimensional spatial networks and their continuum limit}

\renewcommand{\thefootnote}{\fnsymbol{footnote}}%
\author{
Adel Mehrpooya$^\text{a,b}$, Vivien J.~Challis$^\text{a,b}$ and Pascal R.~Buenzli$^\text{a,b,}$\footnotemark[1]}

\date{\small $^\text{a}$School of Mathematical Sciences, Queensland University of Technology (QUT), Brisbane, Australia\\\small $^\text{a}$Max Planck Queensland Centre for the Materials Science of Extracellular Matrices, Queensland University of Technology (QUT), Brisbane, Australia\\\vskip 1mm \normalsize\today\vspace*{-5mm}}

\maketitle
\begin{abstract}
	The propagation of signalling molecules within cellular networks is affected by network topology, but also by the spatial arrangement of cells in the networks. Understanding the collective reaction--diffusion behaviour in space of signals propagating through cellular networks is an important consideration, for example for regenerative signals that convey positional information. In this work, we consider stochastic and deterministic  versions  of random walk models of signalling molecules propagating and reacting within one-dimensional spatial networks with arbitrary node placement and connectivity. By taking a continuum limit of the random walk models, we derive an inhomogeneous reaction--diffusion--advection equation, where diffusivity and advective velocity depend on local node density and connectivity within the network. Our results show that large spatial variations of molecule concentrations can be induced by heterogeneous node distributions.  Furthermore, we find that noise within the stochastic random walk model is directly influenced by node density. We apply our models to consider signal propagation within the osteocyte network of bone, where signals propagating to the bone surface regulate bone formation and resorption processes. We investigate signal-to-noise ratios for different damage detection scenarios and show that the location of perturbations to the network can be detected by signals received at the network boundaries.

\end{abstract}

\protect\footnotetext[1]{Corresponding author. Email address: \texttt{pascal.buenzli@qut.edu.au}}%
\renewcommand{\thefootnote}{\arabic{footnote}}%

\section{Introduction}
Biological tissues are composed of complex arrangements of cells in space that enable them to perform a variety of functions. To understand how these functions emerge from cell--cell interactions taking place in these complex arrangements, new quantitative approaches based on networks have been developed~\cite{graner-etal-2008,fischer-bassel-kollmannsberger-2023,strauss-etal-2022,vicente-munuera-etal-2020}. These networks represent cell--cell communication pathways due for example to the adjacency of cells in plants and  confluent epithelial cell layers~\cite{escudero-etal-2011,jackson-etal-bassel-2017,jackson-etal-bassel-2019}, or due to branching cellular dendrites connecting near and far cells, such as in the neural network in the brain~\cite{sporns2013structure,sporns2005human} and in the osteocyte network in bone~\cite{kerschnitzki-etal-2013,mader-etal-2013,buenzli-sims-2015,kollmannsberger-etal-2017}. Distributed networks also arise in cardiovascular, respiratory, plant vascular and root systems~\cite{banavar-maritan-rinaldo-1999}. In all these biological systems, understanding how network architecture influences transport properties of signals propagating through these networks is important to assess tissue functional behaviour~\cite{strogatz-2001}.  For example, transport properties may affect the efficiency of nutrient delivery and waste product removal in surrounding tissues. In plants, signalling through the cell-to-cell communication network regulates growth, patterning, and defense mechanisms~\cite{maule-etal-2012,jackson-etal-bassel-2019}. 

Random walks are an effective way to describe stochastic reaction--diffusion processes occurring on networks~\cite{masuda2017random,porter-gleeson-2016}. They have been used to describe disease spread in social networks, the spread of opinions in voting models~\cite{vilone-vespignani-castellano-2002}, and transportation systems~\cite{barrat-barthelemy-2008,barrat-barthelemy-vespignani-2005}. Several common reaction--diffusion continuum models, such as Fisher--KPP and activator--inhibitor models, are generalised to networks using discrete Laplacians to investigate signal propagation~\cite{merris-1994,hens-etal-2019,falco-2022,putra-etal-goriely-2023} and spatio-temporal patterns induced by dynamic instabilities~\cite{othmer-scriven-1971,othmer-scriven-1974,horsthemke-lam-moore-2004,nakao-mikhailov-2010,tambyah-etal-2021}.

Most studies of signal propagation in networks focus on the impact of the topological graph structure. However, many biological networks are embedded in physical space~\cite{barthelemy-2018}. The function of some of these networks relies on signals transmitting positional information, which depends not only on network topology, but also on the physical location of the network's nodes and edges. The osteocyte network in bone, for example, senses local mechanical deformation of bone tissue and detects micro-damage to direct bone adaptation and bone regeneration where needed~\cite{bonewald-2011,manolagas-parfitt-2013,jilka-noble-weinstein-2013}. This occurs through signalling molecules propagating through the osteocyte network, or through the lacuno-canalicular network in which osteocytes reside, from within bone tissue to the bone surface. Recent mathematical models have simulated fluid flow in lacuno-canalicular networks of mouse tibia using circuit theory on weighted graphs~\cite{vanTol-etal-2020a,vanTol-etal-2020b,spielman-2010}.  While these simulations take into account precise network information from micro-CT scans, they provide limited mathematical understanding of the effect of network architecture on average signal transmission at the tissue scale. They also do not take into account the effect of chemical reactions such as creation and degeneration of signalling molecules and the importance of stochasticity in these processes. 

In this paper, we investigate random walk models of signal propagation in one-dimensional spatial networks where node placement and connections between nodes is arbitrary  (Section~\ref{sec:discrete}). We assess the collective reaction-diffusion behaviour of these models by deriving a continuum limit in which the physical spacing between the nodes is small (Section~\ref{sec:continuum}). The theory of random walks on regular lattices is well-developed and known to be described by reaction--diffusion phenomena~\cite{codling2008random,weiss1994,van1992stochastic,falco-2022}.
Here, we show that on irregular networks, the continuum limit leads to an inhomogenous reaction--diffusion--advection partial differential equation for molecule concentration. The diffusivity and advective velocity in this equation depend explicitly on local node density and connectivity.  We also formulate an equivalent deterministic position-jump model that captures the average behaviour of the stochastic random walk model and provides fast and accurate solutions to the continuum equation~\cite{angstmann-henry-etal-2019} (Section~\ref{sec:discrete}). To explore the role of the spatial arrangement of nodes, we perform numerical simulations of random walks on a series of irregular spatial networks and compare these results with numerical solutions of the reaction--diffusion--advection equation  (Section~\ref{sec:results}). These comparisons emphasise the importance of the spatial structure of networks for quantities related to signal propagation, such as transport and signal-to-noise ratios. Finally, we provide an application of these models to the osteocyte network in bone (Section~\ref{sec:osteocytes}). The reaction--diffusion--advection equation is of particular interest in this system to capture coarse-grained behaviour of signal propagation through this dense, large spatial network~\cite{buenzli-sims-2015}.

\section{Methods}
 We first present the stochastic and deterministic variants of the random walk model in Section~\ref{sec:discrete}. The evolution of average node occupancy is governed by Eq.~\eqref{evo-N}, and a relationship between the variance of the molecular concentration and the node density is established in Eq.~\eqref{var-n}. We then derive a continuum limit of the deterministic model in Section~\ref{sec:continuum}, which leads to a reaction--diffusion--advection equation (Eq.~\eqref{evo-n}) with a diffusivity and an advective velocity that depend explicitly on the network's node density and connectivity (Eqs~\eqref{D}--\eqref{v}). 

\subsection{{Stochastic and  deterministic  discrete models}}\label{sec:discrete} 

\begin{figure}	
	\centering
	\includegraphics[scale=0.33]{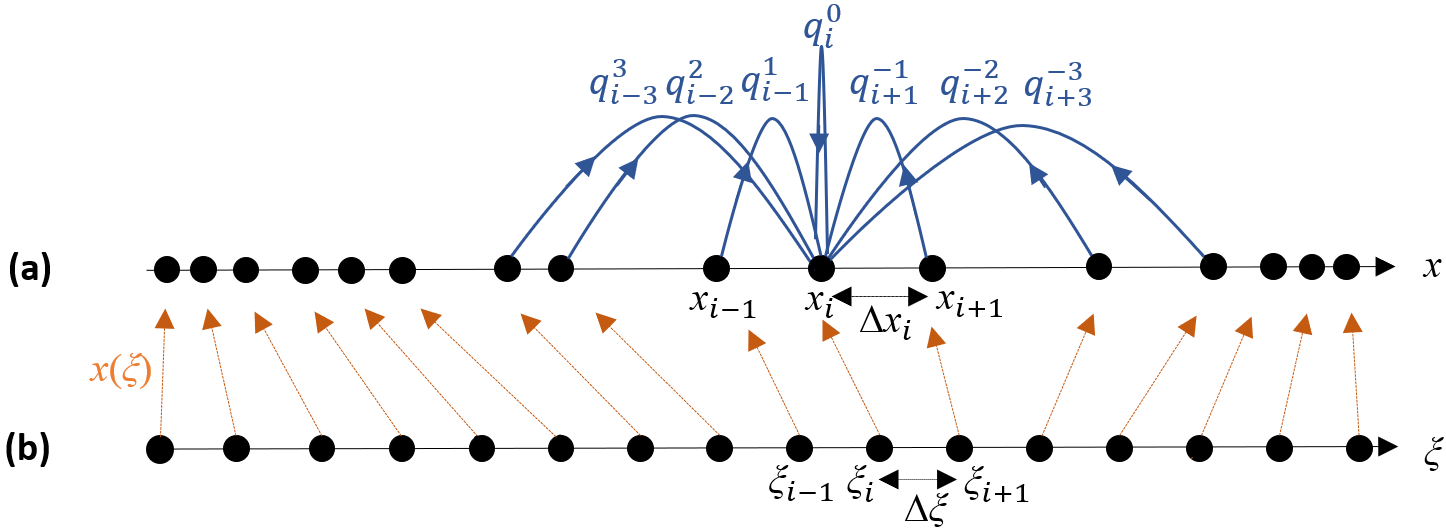}
	\caption{(a) Schematic of a random walk on irregularly spaced nodes $x_i$, with only the incoming connections to node $i$ displayed. (b) The regular network of nodes $\xi_i$ is mapped to the $x_i$ network by $x(\xi)$.}
	\label{fig-axes-schematic}
\end{figure}

We consider a general discrete-time random walk model for the propagation of signalling molecules in a one-dimensional spatial network. The network's nodes are distributed arbitrarily along the $x$-axis at positions $x_i$, $i = 1, \ldots, K$, that are ordered by increasing value (Fig.~\ref{fig-axes-schematic}a). Each node $i$ is assumed to hold a number of signalling molecules $N_i(t)$ at time $t$. We define the territory of node $i$, $\Deltaup x_i$, as half the distance to node $i - 1$ plus half the distance to node $i + 1$:
\begin{align}\label{eq02}
	\Deltaup x_i = \dfrac{1}{2}(x_i - x_{i-1}) + \dfrac{1}{2}(x_{i+1} - x_{i}) = \dfrac{1}{2}(x_{i+1} - x_{i-1}).
\end{align}
This spatial network has a node density at $x_i$ given by
\begin{align}\label{rho_i}
	\rho_i = \frac{1}{\Deltaup x_i}.
\end{align}
The concentration $n_i(t)$ of molecules at $x_i$ at time $t$ is defined as
\begin{align}\label{n_i}
	n_i(t) = \dfrac{N_i(t)}{\Deltaup x_i} = \rho_i N_i(t).
\end{align}
Each node $i$ is assumed to be connected to an arbitrary number of nodes of the network among all available nodes $j=1,\ldots,K$. We introduce $q_i^s$ as the probability for a signalling molecule to make a jump of size $s$ from node $i$ during a time increment $\Deltaup t$, where $s$ can be positive, negative, or zero, i.e., $q_i^s$ is the probability for the signalling molecule to jump from node $i$ to node $j=i+s$ during $\Deltaup t$. The network is assumed to be directed, such that the probability of jumping from node $i$ to node $j$ may not necessarily equal the probability  of jumping from node $j$ to node $i$.

Signalling molecules may also undergo chemical reactions during the same time increment $\Deltaup t$. We assume that they may be created at node $i$ with a probability per unit time $C(N_i)$, and eliminated at node $i$ with a probability per unit time $E(N_i)$. These rates depend on the number of molecules present at the node and may also vary explicitly with position and time. For ease of notation we omit these additional function arguments. The net chemical reaction rate at node $i$ is
\begin{align}
	F(N_i) = C(N_i) - E(N_i).
\end{align}

The evolution of the stochastic model is determined by formulating the balance of molecules at node $i$ from time $t$ to time $t+\Deltaup t$. The number of molecules at node $i$ at time $t+\Deltaup t$ either comes from molecules making a jump of size $s$ from node $i-s$, or is due to creation or elimination of molecules according to the chemical reaction rate. Jump size $s$ is an offset of node indices and may be a positive or negative integer, or zero if molecules remain at node $i$. On average, the number of molecules at node $i$ is expected to evolve according to
\begin{align}\label{evo-N}
	N_i(t+\Deltaup t) = \sum_s q_{i-s}^s\,N_{i-s}(t) + F\big(N_i(t)\big)\Deltaup t,  
\end{align}
where the sum is taken over all possible jump sizes $s$ leading to node $i$, which depends on the connectivity of node $i$ (Fig.~\ref{fig-axes-schematic}a)~\cite{weiss1994,van1992stochastic}. Alternatively, one may consider the network to be complete (every node is connected to every node), but jump probabilities are zero where nodes are not connected. Jumps conserve molecules, so that the sum of jump probabilities for molecules leaving internal nodes must satisfy:
\begin{align}\label{sum-q}
	\sum_s q_i^s = 1, \qquad i=2,\ldots,K-1. 
\end{align}

At the boundaries of the network we assume absorbing boundary conditions, so that any incoming molecules at nodes $i=1$ and $i=K$ are removed from the system, resulting in
\begin{align}\label{BCs-N}
	N_1(t) = 0,\qquad N_K(t) = 0 \qquad \forall\ t.
\end{align}
In practice, we keep a record of the number of molecules $N_-(t)$ and $N_+(t)$ reaching the boundary nodes $i=1$ and $i=K$, and calculate the flux of signalling molecules leaving the network at the left and right boundaries as
\begin{align}
	J_-(t) = \dfrac{N_-(t)-N_-(t - \Deltaup t)}{\Deltaup t}, \quad \text{ and } \quad J_+(t) = \dfrac{N_+(t)-N_+(t - \Deltaup t)}{\Deltaup t},\label{flux-discrete}
\end{align}
respectively. Both of these fluxes are non-negative. 

The \emph{stochastic model} presented above provides evolution rules for the stochastic variable $N_i^\text{stoch}(t)$ representing the number of signalling molecules at node $i$ at time $t$  (see \emph{Numerical simulations} below). The stochastic model is associated with a master equation \cite{barrat-barthelemy-2008,codling2008random,van1992stochastic,weiss1994,masuda2017random} that describes the evolution of the probability $P_i(N,t)$ of having $N$ molecules at node $i$ at time $t$.  In this work, we are particularly interested in describing the average behaviour of signal propagation in space in the continuum limit, which is related to $\langle N_i^\text{stoch}(t)\rangle$ calculated as the first moment of $P_i(N,t)$. A closed-form evolution law for the average occupancy of molecules $\langle N_i^\text{stoch}(t)\rangle$ can only be found in general when reaction rates are linear or under a mean-field approximation~\cite{van1992stochastic,erban-chapman-2009,erban-chapman-maini-2007}. Instead of using the master equation to derive an evolution law for the average, here we propose a deterministic, discrete version of the random walk model and consider its continuum limit. Both the deterministic discrete model and its continuum limit capture the average behaviour of the stochastic random walk model.  The deterministic variant of the stochastic model is obtained by considering $q_i^s$ to represent the \emph{fraction} of molecules making a jump of size $s$ from node $i$. In this case, Eq.~\eqref{evo-N} provides an evolution rule for a (non-integer) variable $N_i^\text{disc}(t)$ such that $N_i^\text{disc}(t) \approx \big\langle N_i^\text{stoch}(t)\big\rangle$. We refer to this deterministic variant of the model as the \emph{ deterministic  model}.

\subsubsection*{Noise}
The distribution of molecules $N_i^\text{stoch}(t)$ across the network and corresponding concentration $n_i^\text{stoch}(t)$ from Eq.~\eqref{n_i} are random variables, whose variance quantifies the expected variability of signals propagating through the network. When there are no reactions ($C(N)=0$, $E(N)=0$), molecules propagate through the network without interacting with each other. The joint probability $P_{i_1,\ldots,i_{N_\text{tot}}}(t)$ that at time $t$ molecule~1 is at node~$i_1$, molecule~2 is at node~$i_2$, etc., factorises into
$
  P_{i_1,\ldots,i_{N_\text{tot}}}(t) = P_{i_1}(t)P_{i_2}(t)\cdots P_{i_{N_\text{tot}}}(t),
$
where $N_\text{tot}$ is the total number of molecules in the network and $P_{i}(t)$ is the single-molecule probability of being at node~$i$ at time~$t$. The probability distribution of $N_i^\text{stoch}(t)$ is the probability $P_i(N,t)$ of encountering~$N$ molecules at node~$i$ at time~$t$, given by
\begin{align}
  P_i(N,t) &= \text{Prob}\big\{N_i^\text{stoch}(t)=N\big\}\notag
  \\&=\sum_\sigma P_{\sigma(\underbrace{i,i,\ldots,i}_{N\ \text{times}},\displaystyle i_1,i_2,\ldots,i_{N_\text{tot}-N})}(t)\notag
  \\&= {N_\text{tot} \choose N} {P_i(t)}^N \left( 1- P_i(t)\right)^{N_\text{tot}-N},\label{binomial}
\end{align}
where the sum in Eq.~\eqref{binomial} is taken over all $N_\text{tot}!$ permutations $\sigma$ of node indices $i_1,i_2,\ldots,i_{N_\text{tot}}$. Eq.~\eqref{binomial} shows that at each time $t$, $N_i^\text{stoch}(t)$ follows a binomial distribution with parameters $N_\text{tot}$ and $P_i(t)$. The mean and variance of $N_i^\text{stoch}(t)$ are therefore given by~\cite{ross2014first}
\begin{align}
  &\Big\langle N_i^\text{stoch}(t)\Big\rangle = N_\text{tot}\, P_i(t),\label{avg-N}
  \\&\text{Var}\Big(N_i^\text{stoch}(t)\Big) = \Big\langle{N_i^\text{stoch}(t)}^2\Big\rangle - \Big\langle N_i^\text{stoch}(t)\Big\rangle^2 = N_\text{tot}\,P_i(t)\big(1-P_i(t)\big).\label{var-N}
\end{align}
The variance of the concentration of molecules $n_i^\text{stoch}(t)$ is found by using Eq.~\eqref{n_i} and~\eqref{var-N}. We first relate
$P_i(t)$ to the average concentration of molecules along $x$ by Eqs~\eqref{avg-N} and~\eqref{n_i},
\begin{align}\label{P_i}
	P_i(t) = \frac{1}{N_\text{tot}\ \rho_i}\Big\langle n_i^\text{stoch}(t)\Big\rangle.
\end{align}
Substituting this expression in Eq.~\eqref{var-N}, and since we expect that $\big\langle n_i^\text{stoch}(t)\big\rangle \approx n_i^\text{disc}(t)$,
we have
\begin{align}
  \text{Var}\big(n_i^\text{stoch}(t)\big) &= \rho_i^2\ \text{Var}\Big(N_i^\text{stoch}(t)\Big) \notag \\&= \rho_i \Big\langle n_i^\text{stoch}(t)\Big\rangle - \frac{1}{N_\text{tot}} \Big\langle n_i^\text{stoch}(t)\Big\rangle^2 \notag \\ &\approx \rho_i n_i^\text{disc}(t) - \frac{\big(n_i^\text{disc}(t)\big)^2}{N_\text{tot}}. \label{var-n}
\end{align}

\subsubsection*{Numerical simulations}
Numerical simulations of the  deterministic  model are performed by updating the number of signalling molecules $N_i^\text{disc}(t)$ by stepping forward in time according to Eqs~\eqref{evo-N}, \eqref{BCs-N} from an initial condition $N_i^\text{disc}(0)$ for $i=1,\ldots,K$.

Numerical simulations of the stochastic model update the number of signalling molecules $N_i^\text{stoch}(t)$ algorithmically in two steps: a transport step, and a reaction step. These steps can be considered to occur sequentially during the same time increment $\Deltaup t$ under the assumption that jump probabilities are independent of the number of molecules, and consequently, independent of chemical reactions that may occur during $\Deltaup t$~\cite{van1992stochastic}. The transport step introduces an intermediate quantity, $\widetilde{N}_i^{\text{stoch}}$, that represents the number of molecules at node $i$ after transport but before any reaction. This quantity is first set to $\widetilde{N}_i^\text{stoch}=N_i^\text{stoch}(t)$ at all nodes $i=1,\ldots,K$, and updated by accounting for jumps made by the molecules during the time increment $\Deltaup t$ at all internal nodes $i=2,\ldots,K-1$. For each molecule at node $i$, $i=2,\ldots,K-1$, we select a target node $i+s$ with probability $q_i^s$, decrement $\widetilde{N}_i^\text{stoch}$ by one, and increment $\widetilde{N}_{i+s}^\text{stoch}$ by one. Any jump that would take a signalling molecule out of the network ($i+s\geq K$ or $i+s\leq 1$) results in the molecule being removed from the system and contributes to the flux at the boundary.

After this transport step, $\widetilde{N}_i^\text{stoch}$ is used to process the reaction step. The probability of undergoing a creation or elimination reaction during $\Deltaup t$ at node $i$ is \cite{van1992stochastic}
\begin{align}\label{P_reac}
	P_\text{reac}(i,t) = C\big(\widetilde{N}_i^\text{stoch}\big)\Deltaup t + E\big(\widetilde{N}_i^\text{stoch}\big)\Deltaup t.
\end{align}
We assume that $\Deltaup t$ is small enough that $P_\text{reac}(i,t)<1$, and that at most a single reaction occurs at each node within the time increment $\Deltaup t$. If a reaction at node $i$ occurs, then the probability that it is a creation event is
\begin{equation}\label{P_C}
	P_{C}(i,t) = \dfrac{C\big(\widetilde{N}_i^\text{stoch}\big)}{C\big(\widetilde{N}_i^\text{stoch}\big) + E\big(\widetilde{N}_i^\text{stoch}\big)},
\end{equation}
in which case $N_i^\text{stoch}(t+\Deltaup t) = \widetilde{N}_i^\text{stoch}+1$. Conversely, the probability of the reaction being an elimination event is
\begin{equation}\label{P_E}
	P_{E}(i, t) = \dfrac{E\big(\widetilde{N}_i^\text{stoch}\big)}{C\big(\widetilde{N}_i^\text{stoch}\big) + E\big(\widetilde{N}_i^\text{stoch}\big)},
\end{equation} 
in which case $N_i^\text{stoch}(t+\Deltaup t) = \widetilde{N}_i^\text{stoch}-1$.

\subsection{Continuum limit}\label{sec:continuum}
When spacings between nodes of the network and the time increment $\Deltaup t$ are small, the behaviours of the stochastic  and deterministic discrete models  can be approximated by a continuum conservation equation for the concentration of signalling molecules. On regular lattices with even spacings and jumps to neighbouring lattice sites, the derivation of continuum limits of random walks is well established and used as a model of Brownian motion and molecular diffusion~\cite{codling2008random,masuda2017random,van1992stochastic,weiss1994}. Here, we present a derivation of the continuum limit for arbitrary one-dimensional spatial networks in which node positions and connections are arbitrary, based on the evolution equation of the  deterministic  model given by Eq.~\eqref{evo-N}.

To formally take the limit as the spacing between the nodes along the $x$-axis approaches zero while retaining irregular spacings between nodes, we first replace the discrete mapping $i \longmapsto x_i$ by a bijective continuous mapping $\xi\mapsto x(\xi)$ such that
\begin{equation}
	\xi_i \longmapsto x_i = x(\xi_i),
\end{equation}
in which $\xi_i = i\Deltaup \xi$ are evenly spaced coordinates along an axis $\xi$ (Fig.~\ref{fig-axes-schematic}b). The quantity $\Deltaup\xi$ represents a spatial scaling factor, so that the continuum limit is then taken by letting $\Deltaup \xi \to 0$ and $\Deltaup t \to 0$. Since node positions $x_i$ are ordered, we can assume that $x(\xi)$ is a monotone increasing function. With the mapping $x(\xi)$, the territory of node $i$ is approximated by
\begin{equation}\label{gdxi}
	\Deltaup x_i \approx g(\xi_i) \Deltaup \xi,
\end{equation}
where
\begin{align}\label{g}
	g(\xi) = \frac{\mathrm{d} x}{\mathrm{d} \xi} > 0
\end{align}
is the metric associated with $x(\xi)$. We define the continuous local density of nodes at position $x$ by
\begin{align}\label{rho}
	\rho(x) = \frac{1}{g\big(\xi(x)\big)\Deltaup\xi},
\end{align}
where $\xi(x)$ is the inverse function of $x(\xi)$, such that from Eqs~\eqref{rho_i} and~\eqref{gdxi}, $\rho(x_i)\approx \rho_i$.

The distribution of molecules along the $x$-axis governed by Eq.~\eqref{evo-N} is mapped by the inverse function $\xi(x)$ into a corresponding distribution of molecules along the $\xi$-axis (Fig.~\ref{fig-axes-schematic}). We first derive the continuum limit of the evolution equation for the concentration of molecules along the $\xi$-axis, where nodes are evenly spaced, and then map this evolution back onto the $x$-axis via $x(\xi)$. We introduce continuous functions $\eta(\xi,t)$ and $n(x,t)$ representing the concentration of molecules along the $\xi$-axis and along the $x$-axis, respectively, such that
\begin{align}\label{eta-n-N}
  \eta(\xi_i,t) = \frac{N_i(t)}{\Deltaup \xi}, \qquad n(x_i,t) = n_i(t) \approx \frac{N_i(t)}{g(\xi_i)\Deltaup\xi}  = \rho(x_i)N_i(t).
\end{align}
The molecule concentrations along the $x$-axis and along the $\xi$-axis are related by
\begin{align}\label{n-eta}
	n(x, t) = \dfrac{\eta(\xi(x), t)}{g(\xi(x))}.
\end{align}
We also introduce continuous functions of space $q(\xi,s)$ representing the probability for a molecule at position $\xi$ to make a jump of size $s\Deltaup\xi$ along the $\xi$-axis, such that $q(\xi_i,s) = q_i^s$. With these definitions, Eq.~\eqref{evo-N} divided by $\Deltaup \xi$ becomes
\begin{align}\label{evo-eta1}
	\eta(\xi_i, t+\Deltaup t) = \sum_s q(\xi_{i-s}, s)\eta(\xi_{i-s}, t) + \frac{F\big(\eta(\xi_i,t)\Deltaup\xi\big)}{\Deltaup\xi}\Deltaup t.
\end{align}
Expanding Eq.~\eqref{evo-eta1} about $(\xi_i,t)$ to first order in $\Deltaup t$ and second order in $\Deltaup \xi$ using $\xi_{i-s} = \xi_i - s\Deltaup\xi$ gives
\begin{align}\label{evo-eta2}
  \eta(\xi_i,t) + \pd{}{t}\eta(\xi_i,t)\Deltaup t + \Order(\Deltaup t^2) = &\sum_sq(\xi_i,s)\eta(\xi_i,t)-\pd{}{\xi}\sum_sq(\xi_i, s)\, s\ \eta(\xi_i,t)\Deltaup\xi \notag \\&+ \frac{1}{2}\pd{^2}{\xi^2}\sum_sq(\xi_i,s)\,s^2\ \eta(\xi_i,t)\Deltaup\xi^2 + \Order\big(\Deltaup \xi^3\big) \notag \\& + \frac{F\big(\eta(\xi_i,t)\Deltaup\xi\big)}{\Deltaup\xi}\Deltaup t. 
\end{align}
Assuming now that the equation holds for any value of $\xi$, using Eq.~\eqref{sum-q}, and noting that 
\begin{align}\label{jump-moments}
	\avg{s(\xi)} = \sum_s q(\xi,s)\, s, \qquad \big\langle s^2(\xi) \big\rangle = \sum_s q(\xi,s)\,s^2
\end{align}
are the first and second moments of the jump probabilities, respectively, Eq.~\eqref{evo-eta2} becomes
\begin{align}\label{evo-eta3}
  \pd{}{t}\eta(\xi,t) = &- \pd{}{\xi}\left[\frac{\avg{s(\xi)}\Deltaup\xi}{\Deltaup t}\eta(\xi,t) - \pd{}{\xi}\left(\frac{\big\langle s^2(\xi)\big\rangle\Deltaup\xi^2}{2\Deltaup t}\eta(\xi,t)\right)\right] + \frac{F\big(\eta(\xi,t)\Deltaup\xi\big)}{\Deltaup\xi},
\end{align}
where we neglected terms of order $\Order(\Deltaup t)$ and  $\Order(\Deltaup\xi^3/\Deltaup t)$ because they are subdominant under the assumption that the continuum limit $\Deltaup t\to 0$ and $\Deltaup\xi\to 0$ is taken such that the ratio $\Deltaup\xi^2/\Deltaup t$ is constant. We also assume that jump probability biases are proportional to $\Deltaup\xi$ so that $\avg{s(\xi)}\Deltaup\xi/\Deltaup t = \Order\big(\Deltaup\xi^2/\Deltaup t\big)$ and both terms in the square bracket are of order $\Order(1)$ in the continuum limit. The above expansion in $\Deltaup\xi$ is valid provided node degrees are finite, or $q(\xi,s)\to 0$ as $s\to\pm\infty$ uniformly in $\xi$ and sufficiently fast that the moments in Eq.~\eqref{jump-moments} are well-defined. Finally, we define
\begin{align}\label{Dtilde-vtilde}
	\widetilde{D}(\xi) = \frac{\big\langle s^2(\xi)\big\rangle \Deltaup\xi^2}{2\Deltaup t}, \qquad \widetilde{v}(\xi) = \frac{\avg{s(\xi)}\Deltaup\xi}{\Deltaup t} - \pd{}{\xi}\widetilde{D}(\xi),
\end{align}
and expand $\pd{}{\xi}\big(\widetilde{D}(\xi)\eta(\xi,t)\big) = \big(\pd{}{\xi}\widetilde{D}(\xi)\big)\eta(\xi,t) + \widetilde{D}(\xi)\pd{}{\xi}\eta(\xi,t)$ in Eq.~\eqref{evo-eta3} to obtain
\begin{align}\label{evo-eta}
	\pd{}{t}\eta(\xi,t) = -\pd{}{\xi}\left[\widetilde{v}(\xi)\eta(\xi,t) - \widetilde{D}(\xi)\pd{}{\xi}\eta(\xi,t)\right] + \frac{F\big(\eta(\xi,t)\Deltaup\xi\big)}{\Deltaup\xi}.
\end{align}
Equation~\eqref{evo-eta} is a reaction--diffusion--advection equation for the concentration of molecules along the $\xi$-axis, with inhomogenous diffusion $\widetilde{D}(\xi)$ and inhomogenous advective velocity $\widetilde{v}(\xi)$ that depend on node connectivity and jump probabilities via Eqs~\eqref{Dtilde-vtilde} and~\eqref{jump-moments}. We note here that this developement is similar to a Kramers-Moyal expansion  of the master equation that leads to the Fokker-Planck or Smoluchowski equation, a partial differential equation (PDE) that governs the time evolution of the probability density function of molecule distribution in space~\cite{swain1984handbook,van1992stochastic,weiss1994,codling2008random}. The advantage of not considering the master equation is that we can derive a PDE governing the average molecule concentration in the continuum limit more directly. To derive an evolution equation for the average concentration from the master equation would require evaluating the first moment of the master equation, and considering mean-field approximations to remove nonlinearities in the reaction term~\cite{van1992stochastic}. One of the difficulties of this approach is to combine both stochastic reactions and stochastic jumps in the master equation. We refer the interested reader to Refs~\cite{erban-chapman-2009,erban-chapman-maini-2007}, in which such an approach is undertaken. The continuum limit leading to Eq.~\eqref{evo-eta} is thus  expected to represent the average behaviour of the stochastic model under a mean-field approximation where there are no long-range correlations and only weak nonlinearities~\cite{van1992stochastic,gillespie2007stochastic}.  In particular, it may not hold well if the network possesses far connections over distances of the same order as the whole domain. The limit $\Deltaup \xi\to 0$ means that distances between connected nodes remain small with respect to the whole domain of the network. In practice, most spatial networks predominantly connect nodes in close spatial proximity because there is usually a cost associated with building connections.

The evolution of the concentration of molecules $n(x,t)$ along the $x$-axis is found from Eq.~\eqref{evo-eta} by using Eq.~\eqref{n-eta} and noting that $\frac{\partial }{\partial x} = \frac{1}{g}\frac{\partial}{\partial \xi}$. For ease of notation, we now omit space and time arguments, and implicitly assume that all spatial arguments can be either functions of $\xi$ or $x$, as appropriate, via the one-to-one mapping $x(\xi)$. We have
\begin{align}\label{n-eta-dev}
	\dfrac{\partial \eta}{\partial t} = g \dfrac{\partial n}{\partial t}, \quad \text{ and } \quad \dfrac{\partial \eta}{\partial \xi} = \dfrac{\partial n}{\partial \xi} g + n \dfrac{\partial g}{\partial \xi} = g^2 \dfrac{\partial n}{\partial x} + n g \dfrac{\partial g}{\partial x},
\end{align}
so that Eq.~\eqref{evo-eta} becomes
\begin{align}\label{e5}
	\dfrac{\partial n}{\partial t} g = \left(- \dfrac{\partial}{\partial x}\left[\tilde{v} n g - \tilde{D} \left(\dfrac{\partial n}{\partial x} g^2 + n g \dfrac{\partial g}{\partial x}\right)\right] + \dfrac{F(n g \Deltaup \xi)}{g\Deltaup \xi}\right)g.
\end{align}
Therefore, the concentration of molecules $n(x, t)$ obeys the reaction--diffusion--advection equation
\begin{align}\label{evo-n}
	\dfrac{\partial n}{\partial t}
	&= -\dfrac{\partial}{\partial x}\left[ v\,n - D\dfrac{\partial n}{\partial x}\right] + \rho \,F(n/\rho),
\end{align}
where
\begin{align}
  D(x) = g^2\widetilde{D}, \qquad v(x) = g \big[\widetilde{v} - \widetilde{D} \pd{g}{x}\big] \label{D-v}
\end{align}
are a diffusivity and advective velocity in $x$-space, respectively, such that the total flux of signalling molecules at location $x$ and time $t$ is $J(x,t) = v n - D\pd{n}{x}$. The space dependence of $D(x)$ and $v(x)$ arises due to the network's varying node density via the metric~$g$, and due to the node connectivity via $\widetilde D$ and $\widetilde v$. Using Eqs~\eqref{Dtilde-vtilde},~\eqref{jump-moments}, and the correspondence between continuous and discrete jump probabilities, we can elucidate how the local diffusivity $D(x_i)$ and advective velocity $v(x_i)$ at node $i$ depend on the connections of node $i$ with other nodes and the local node density $\rho(x_i)$ as
\begin{align}
  &D(x_i) = \frac{\sum_s q_i^{s} s^2}{2\Deltaup t\,\rho^2(x_i)},\label{D}
  \\&v(x_i) = \frac{\sum_s q_i^{s} s}{\Deltaup t\,\rho(x_i)} - \dfrac{1}{\rho(x_i)}\pd{(\rho D)}{x}(x_i).\label{v}
\end{align}
Equations~\eqref{evo-n}, \eqref{D}, and \eqref{v} provide direct relationships between macro-scale reaction--diffusion--advection properties of signalling molecules along $x$ and local network topography, which includes both placement in space of nodes of the network via the node density $\rho(x)$, and node connectivity via the first and second moments of jump probabilities.

The absorbing boundary conditions~in Eq.~\eqref{BCs-N} assumed for the stochastic and  deterministic  discrete models carry over in the continuum limit, such that
\begin{align}
  &\eta(\xi_1,t) = \eta(\xi_K,t) = 0, \quad\forall t,
  \\&n(x_1,t) = n(x_K,t) = 0, \quad \forall t. \label{BCs-n}
\end{align}
Initial conditions assumed in the stochastic and  deterministic  discrete models determine initial conditions $\eta(\xi,0)$ and $n(x,0)$ for the concentration of molecules via Eq.~\eqref{eta-n-N}.

The flux of molecules $J_\pm(t)$ reaching the boundaries of the domain at time $t$ in the continuum limit has no contribution from the advective flux $vn$ due to the boundary conditions~\eqref{BCs-n}. It is due to the diffusive part of the flux only, i.e.,
\begin{align}\label{J-cont}
	J_-(t) = D(x_1)\pd{n(x_1,t)}{x}, \qquad J_+(t) = -D(x_K)\pd{n(x_K,t)}{x}.
\end{align}
The sign of the flux $J_-(x,t)$ at the left boundary of the continuum model is chosen to ensure consistency with the non-negative flux at the left boundary in the  stochastic and determinisitc discrete  models, see Eqs~\eqref{flux-discrete}.

When all the jump probabilities $q_{i}^{s}$ are set to zero except for $s=0, \pm 1, \pm 2$, we obtain jumps to nearest neighbours and next-nearest neighbours only. In this case, using the fact that \mbox{$q_i^0=1-q_i^{-1}-q_i^1-q_i^{-2}-q_i^2$}, we have
\begin{align}
	&\widetilde{D} = \dfrac{(\Deltaup \xi)^2}{2\Deltaup t}[q_{i}^{1} + q_{i}^{-1} + 4(q_{i}^{2} + q_{i}^{-2})], \label{Dtilde-nn-nnn}
	\\&\widetilde{v} = \left(\dfrac{\Deltaup \xi}{\Deltaup t}[q_{i}^{1} - q_{i}^{-1} + 2(q_{i}^{2} - q_{i}^{-2})] - \dfrac{\partial}{\partial \xi}\widetilde{D}(\xi) \right).\label{vtilde-nn-nnn}
\end{align}
To ensure that advection is of the same order as diffusion in the continuum limit, the jump probabilities must be chosen such that the following biases are proportional to $\Deltaup \xi$:
\begin{align*}
	q_i^{1}-q_i^{-1} = \Order(\Deltaup \xi), \qquad q_i^{2}-q_i^{-2} = \Order(\Deltaup \xi).
\end{align*}

\subsubsection*{Numerical discretisation}
To solve the reaction--diffusion--advection Eq.~\eqref{evo-n} numerically, we use a simple finite difference scheme based on a regular discretisation of $x$, $x_\alpha = x_1 + \alpha\updelta x$, $\alpha = 0, 1, \ldots, M$, $\deltaup x = \frac{x_K-x_1}{M}$, and a regular discretisation of $t$, $t_\gamma = \gamma\updelta t$, $\gamma=0,1,2,\ldots$. The numerical solution $n(x_\alpha,t_\gamma)$ is then stepped in time using a simple explicit forward Euler scheme with a first-order upwind approximation of the advective term $-\pd{}{x}\big(n(x_\alpha,t_\gamma)v(x_\alpha)\big)$, and a second-order central difference approximation of the diffusion term $\pd{}{x}\big(D(x_\alpha)\pd{n(x_\alpha,t_\gamma)}{x}\big)$~\cite{leveque2007finite}. We ensure that $\updelta t$ and $\updelta x$ fulfill the Courant-Friedrichs-Lewy condition
\begin{align}
  \updelta t \leq \min{ \left( \frac{\updelta x}{{\displaystyle\max_x}\{|v(x)|\}}, \frac{\updelta x^2}{2\,{\displaystyle\max_x}\{D(x)\}} \right) }.
\end{align}

\section{Results and discussion}\label{sec:results}

We present a range of numerical results for the stochastic and  deterministic  discrete models and their continuum limit. We first validate the models on a regular network with space-independent jump probabilities prior to increasing the complexity of the considered numerical examples.  In Sections~\ref{sec:space-indep-jumps} and~\ref{sec:space-dep}, space and time units are  considered to be  arbitrary and omitted in the numerical values presented. In Section~\ref{sec:osteocytes}, our model is applied to signalling in the osteocyte network and we provide relevant space and time scales.

\subsection{Space-independent jump probabilities}\label{sec:space-indep-jumps}
  
In this section we assume that every network node $x_i$, $i=1,\dots,K$ is connected to its immediate neighbours only and that jump probabilities are space-independent. Therefore, $q_i^{-1} = q^{-1}$, $q_i^0=q^0$ and $q_i^1 = q^1$ for $i=2, \dots, K-1$, with all remaining $q_i^s=0$. In addition, we assume that $K$ is odd and the network is symmetrical about the central node $\tilde{i}=\frac{K+1}{2}$ at $x_{(K+1)/2}=0$. All molecules are initialised at the central node of the network at time $t=0$:
\begin{equation}\label{init-cond}
  N_{i}(0) = N_\textrm{tot} \deltaup_{i\tilde{i}} \quad \text{ for } \quad i = 1, \ldots, K, 
\end{equation}
where $N_{\textrm{tot}}$ is the number of initial molecules and $\deltaup_{i\tilde{i}}$ is the Kronecker delta. These initial conditions allow us to consider the response of the network to an impulse signal generated at the centre of the network at time $t=0$. 
In the continuum limit in which $\Deltaup \xi \rightarrow 0$, the initial condition of the reaction--diffusion--advection equation given by Eq.~\eqref{init-cond} becomes a Dirac delta function: 
\begin{equation}\label{limit-init-cond} 
  n(x, 0) = N_\textrm{tot} \updelta(x).
\end{equation}
For the finite difference scheme these initial conditions become  ${N_\mathrm{tot}}/{\updelta x}$ at the discretisation point at $x=0$, and zero elsewhere, where $\updelta x$ is the spatial discretisation and the number of discretisation intervals $M$ is even.

The above assumptions result in constant advection $\widetilde{v}$ and constant diffusion $\widetilde{D}$ in $\xi$-space. In the case that there are no reactions ($F = 0$), Eq.~\eqref{evo-eta} simplifies to a space-homogeneous diffusion-advection equation for $\eta(\xi,t)$. The response to an impulse signal of $N_\mathrm{tot}$ molecules in the centre of an infinite domain in $\xi$-space is therefore given by the fundamental  Gaussian  solution 
\begin{equation*}
  \eta_\infty(\xi,t) = \dfrac{N_{\mathrm{tot}}}{\sqrt{4 \pi \widetilde{D} t}} \exp(-\dfrac{(\xi - \widetilde{v}t)^2}{4 \widetilde{D} t}), \qquad \xi \in (-\infty,\infty).
\end{equation*}
This gives the following infinite-space exact solution $n_{\infty}(x,t)$ for Eq.~\eqref{evo-n} when nodes are positioned arbitrarily, the jump probabilities are space-independent, and there are no reactions:
\begin{equation} \label{n-infty}
  n_{\infty}(x,t) = \dfrac{N_{\mathrm{tot}}}{g(x)\sqrt{4 \pi \widetilde{D} t}} \exp(-\dfrac{(\xi(x) - \widetilde{v}t)^2}{4 \widetilde{D} t}). 
\end{equation}

\begin{figure}
	\centering	
	\includegraphics[scale=0.324]{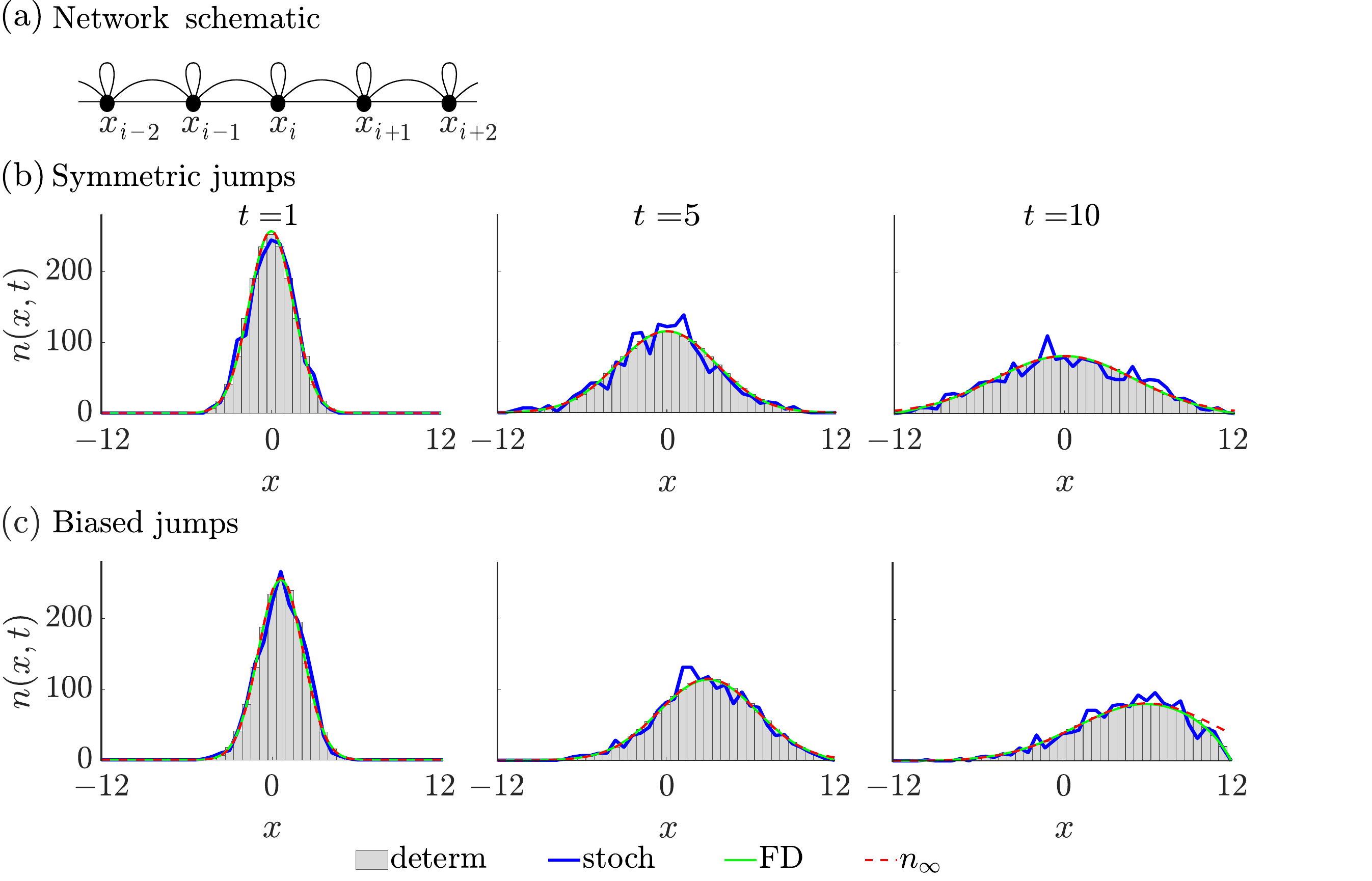}
  \caption{ (a)  Regular network, $x(\xi) = \xi$, $\rho(x)=1$;  (b),(c)  Concentration of molecules corresponding to the  deterministic  (gray), stochastic (blue), finite difference (FD, green), and infinite-space exact (red) solutions at times $t=1$, $t=5$ and $t=10$.  (b): Symmetric jumps ($a=0$) with $q_i^0 = 1/3$, $q_i^{-1} =1/3$, and $q_i^{1}= 1/3$ for internal nodes.  (c): Biased jumps ($a=1/12$) with $q_i^0 = \frac{1}{3}$, $q_i^{-1} = \frac{1}{3} - \frac{5}{100}$, and $q_i^{+1} = \frac{1}{3} + \frac{5}{100}$ for internal nodes. Parameters: $N_\textrm{tot}=1000$, $K=41$, $x_1=-12$, $x_K=12$,  $\Deltaup t = 0.1$, $\Deltaup \xi = \frac{x_K-x_1}{K - 1} = 0.6$, $M = 300$, $\updelta t = 0.0025$, and $\updelta x = \frac{x_K-x_1}{M} = 0.08$. The diffusivity $D = 1.2$ for both cases.}	
	\label{fig-regular}
\end{figure}

We first consider a regular network with 
\begin{equation*}
  x(\xi) = \xi, \quad \xi(x)=x, \quad \text{ and } \quad g(x) = 1.
\end{equation*}
We set jump probabilities as $q_i^0 = 1/3$, $q_i^{-1} =1/3 - a \Deltaup \xi$, and $q_i^{1}= 1/3+a \Deltaup \xi$ for $i = 2, \dots, K-1$, with all other $q_i^s=0$. The parameter $a$ biases jumps to the right (or left) when positive (or negative), or jumps are symmetrical (unbiased) when $a=0$. In this case the advection velocity $v$ and diffusivity $D$ are
\begin{equation*}
  v = \dfrac{(\Deltaup \xi)^2}{\Deltaup t} a \qquad \textrm{and} \qquad D = \dfrac{(\Deltaup \xi)^2}{3\Deltaup t},
\end{equation*}
and we emphasise that they are of the same order in $\Deltaup \xi$ and $\Deltaup t$. 
Figure \ref{fig-regular} presents the concentration of molecules for the stochastic,  deterministic,  and continuum models for the two cases of unbiased and biased jumps. There is an excellent match between the  deterministic  discrete model and finite difference solution to the continuum equation, with variability of the stochastic model around these solutions. The infinite-space exact solution $n_\infty$ matches the finite difference numerical solution until molecules begin reaching the boundaries and start being removed from the finite system. As expected, biased jumps result in molecules on average moving to the right.

\begin{figure}[t]
	\centering
	\includegraphics[scale=0.324]{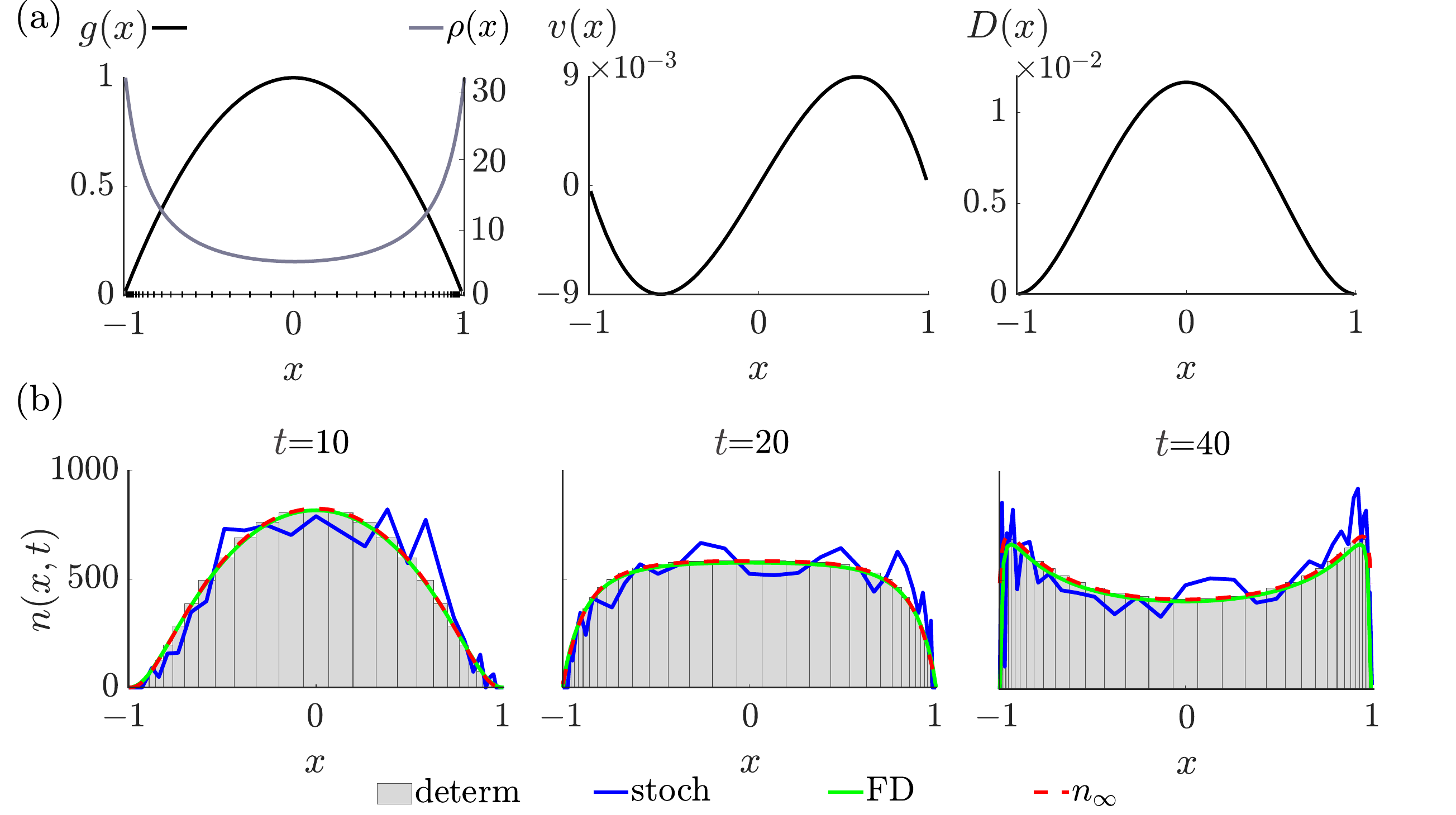}
	\caption{Network with increasing node density, $x(\xi) = \tanh(\xi)$: (a) Metric, density, advection and diffusion functions for this example.  Network node positions are marked on the $x$-axis.  (b) Concentration of molecules corresponding to the  deterministic  (gray), stochastic (blue), finite difference (FD, green), and infinite-space exact (red) solutions at times $t=10$, $t=20$ and $t=40$. 
 Parameters: $q_i^0  = q_i^{-1} = q_i^{1}= \frac{1}{3}$ for internal nodes, $N_\textrm{tot}=1000$, $K=41$, $x_1=-0.99$, $x_K=0.99$, $\Deltaup t = 0.5$, $M = 98$, $\updelta t = 0.0167$.}	
	\label{fig-tanh}
\end{figure}

With the models validated on a regular network, we now consider a network with increasing node density away from the origin. We choose 
\begin{equation}
  x(\xi) = \tanh(\xi),  \qquad \xi(x) = \dfrac{1}{2}\log \dfrac{1+x}{1-x},\qquad \textrm{and} 
  \qquad g(x) = 1 - x^2, 
\end{equation}
where $x \in (-0.99,0.99)$ so that $g(x) > 0$.  
We assume constant and symmetric jump probabilities, $q_i^0  = q_i^{-1} = q_i^{1}= \frac{1}{3}$  for $i = 2, \dots, K-1$. The expressions for the advection $v(x)$ and the diffusion $D(x)$ become, from Eqs~\eqref{D-v},
\begin{equation}
  v(x) = \dfrac{2\Deltaup \xi^2}{3\Deltaup t}x(1 - x^2) \qquad \textrm{and} \qquad 
	D(x) = \dfrac{\Deltaup \xi^2}{3\Deltaup t}(1 - x^2)^2. 
\end{equation}
Figure~\ref{fig-tanh}a illustrates the metric, node density, advection, and diffusion functions for this network.  
Figure \ref{fig-tanh}b shows results for the evolution of concentration of signalling molecules within this network starting from the impulse initial condition in Eqs~\eqref{init-cond},~\eqref{limit-init-cond}. There is again good agreement between the models, with the exception of the infinite-space exact solution $n_\infty$ that does not satisfy the zero Dirichlet boundary conditions on the finite domain. The time snapshots clearly demonstrate how the molecule concentration increases in regions with higher node density. This is expected (see Eqs~\eqref{eta-n-N} and \eqref{n-infty}), and results in an increase in signal strength as the impulse propagates away from the centre of the network. This phenomenon and its relevance to the osteocyte network in bone is explored in more detail in Section~\ref{sec:osteocytes} below.

\begin{figure}[t]
	\centering
	\includegraphics[scale=0.322]{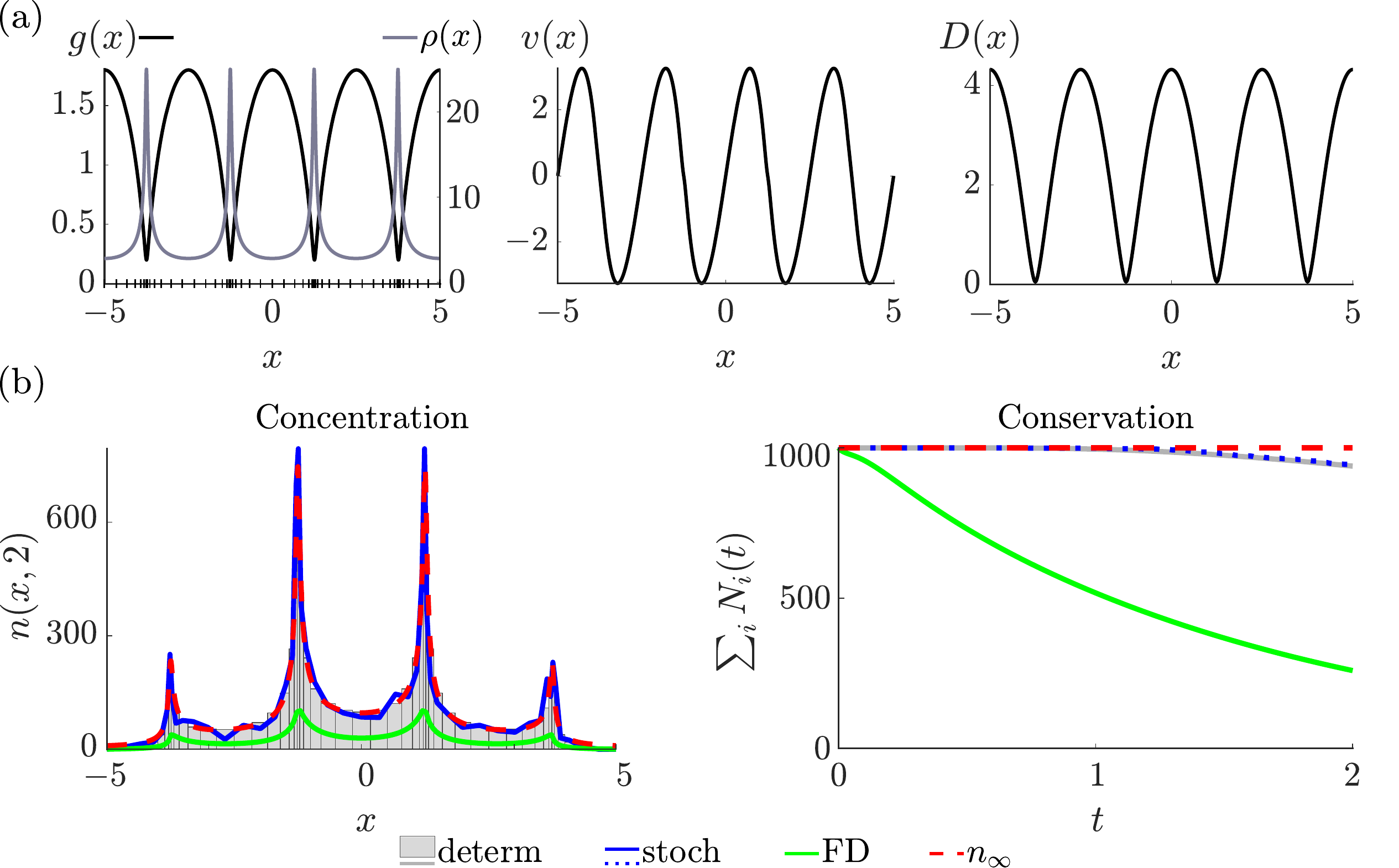}
	\caption{Network with periodic node density, $x(\xi) = \xi + A\sin(k\xi)$: (a) Metric, density, advection and diffusion functions.  Network node positions are marked on the $x$-axis.  (b) Concentration of  molecules corresponding to the  deterministic  (gray), stochastic (blue), finite deference (FD, green), and infinite-space exact (red) solutions at $t = 2$, and conservation of molecules for these models over time. Parameters: $q_i^0  = q_i^{-1} = q_i^{1}= \frac{1}{3}$ for internal nodes, $N_\textrm{tot}=1000$, $K=51$, $x_1=-5$, $x_K=5$, $k = \frac{4\pi}{5}$, $A = \frac{8}{10k}$, $\Deltaup t = 0.01$,  $M = 1200$, $\updelta t = 8\times 10^{-6}$.}
	\label{fig-sin}
\end{figure}

We now consider a network where the node density changes periodically. We choose 
\begin{equation} \label{periodic}
  x(\xi) = \xi + A\sin(k\xi), \qquad 
	g(x) = 1 + Ak\cos(k \xi(x)),
\end{equation}
where $A > 0$ and $k > 0$, and $\xi(x)$ can be calculated by inverting $x(\xi)$ numerically. For the mapping $x(\xi)$ to be bijective, we require $A k < 1$. 
As above, $q_i^0  = q_i^{-1} = q_i^{1}= \frac{1}{3}$  for $i = 2, \dots, K-1$. 
Figure~\ref{fig-sin}a illustrates the functions $g(x)$ and $\rho(x)$ as well as the advection and diffusion functions determined numerically from the following expressions derived from Eqs~\eqref{D-v}:
\begin{equation}\label{eq040}
	v(x) = \dfrac{Ak^2\Deltaup \xi^2}{3\Deltaup t}\sin(k \xi(x)),
\end{equation}
and
\begin{equation}\label{eq041}
	D(x) = \dfrac{\Deltaup \xi^2}{3\Deltaup t}( 1 + Ak\cos(k \xi(x)) )^2. 
\end{equation}
Figure~\ref{fig-sin}b presents a time snapshot of the molecule concentration of the stochastic,  deterministic, finite-difference, and infinite-space solutions, and also presents the total number of molecules for each model over time. As in Fig.~\ref{fig-tanh}, we see the influence of the node density on the signal concentration. Figure \ref{fig-sin}b shows that the finite difference method does not accurately solve the differential equation for this periodic network. This is due to the metric $g(x)$ being small away from the boundaries of the computational domain, which results in a nearly singular reaction--diffusion--advection equation and consequently a significant numerical loss of molecules in this non-conservative finite-difference discretisation scheme. A more sophisticated, conservative numerical scheme could address this issue. However, we emphasise that the  deterministic  discrete model is a fast and effective approach for solving the reaction--diffusion--advection equation that is not affected by small values of $g(x)$ and that remains exactly conservative. The small loss of molecules seen in Figure~\ref{fig-sin}b in the  stochastic and deterministic discrete  models is due to molecules exiting the network at its left and right boundaries.

\subsection{Space-dependent jump probabilities}\label{sec:space-dep}
In this section we consider space-dependent jump probabilities and determine how these probabilities can be selected to give rise to given diffusivity and advection in the continuum limit. This selection is not unique in general so for simplicity, we restrict to nearest neighbour jumps so that only $q_i^{-1}$, $q_i^1$, and $q_i^0=1-q_i^{-1}-q_i^1$ are non-zero. In this case, we can rearrange Eqs~\eqref{D}, and~\eqref{v} to obtain space-dependent jump probabilities $q_i^{-1}$, $q_i^1$, and $q_i^0$ as functions of $D$ and $v$:
\begin{align}
	q_i^{-1} &= \frac{\Deltaup t}{(g \Deltaup \xi)^2} \left(D - \dfrac{\Deltaup \xi}{2}(v g - D g_x + D_{x}g^2)\right),\label{qim1}
  \\q_i^{1} &= \frac{\Deltaup t}{(g \Deltaup \xi)^2} \left(D + \dfrac{\Deltaup \xi}{2}(v g - D g_x + D_{x}g^2)\right),\label{qip1}
  \\q_i^0 &= 1 - \dfrac{2D \Deltaup t}{(g \Deltaup \xi)^2},\label{qi0}
\end{align}
where the right-hand sides are evaluated at $x_i$. Appropriate bounds on $\Deltaup t$ and $\Deltaup \xi$ are required such that all the jump probabilities are non-negative.

For illustration, we solve the spatially homogeneous diffusion equation on an irregular grid using these expressions for jump probabilities. Assuming zero advection ($v=0$) and constant diffusivity ($D=D_{0}$), Eqs~\eqref{qim1}--\eqref{qi0} become
\begin{align}\label{space-dep} 
  q_i^{-1} = \frac{D_0\Deltaup t}{(g \Deltaup \xi)^2} \left(1 + \dfrac{\Deltaup \xi}{2} \pd{g}{x}\right), \quad
	q_i^1 = \frac{D_0\Deltaup t}{(g \Deltaup \xi)^2} \left(1 - \dfrac{\Deltaup \xi}{2} \pd{g}{x}\right) \quad \text{ and }\quad 	q_i^0 = 1 - \dfrac{2D_0 \Deltaup t}{(g \Deltaup \xi)^2}.
\end{align}
For the periodic network given above in Eq.~\eqref{periodic}, the jump probabilities that give the diffusion equation with diffusivity $D_0$ in the continuum limit are
\begin{align}
q_i^{-1} &= \dfrac{D_0\Deltaup t}{((1 + Ak\cos(k \xi(x_i))) \Deltaup \xi)^2} \left(1 - \dfrac{Ak^2\sin(k \xi(x_i))\Deltaup \xi}{2(1 + Ak\cos(k \xi(x_i)))}\right),\notag
  \\	q_i^1 &= \dfrac{D_0\Deltaup t}{((1 + Ak\cos(k \xi(x_i))) \Deltaup \xi)^2} \left(1 + \dfrac{Ak^2\sin(k \xi(x_i))\Deltaup \xi}{2(1 + Ak\cos(k \xi(x_i)))}\right),\notag
	\\	q_i^0 &= 1 - \dfrac{2D_0 \Deltaup t}{((1 + Ak\cos(k \xi(x_i))) \Deltaup \xi)^2}.\label{q-const-diff}
\end{align}
We define the network over the interval $[x_{1}, x_{K}] = [-L, L]$ and continue to use absorbing boundary conditions as in Eq.~\eqref{BCs-n}. We use the initial molecule concentration
\begin{align}
  n(x, 0) = N_{\mathrm{tot}}\frac{\pi}{4L}\cos(\frac{\pi x}{2L}),
\end{align}
and round $N_i(0)= n(x_i,0)\Deltaup x_i$ to the nearest whole molecule at each node $x_i$ for the stochastic model. This choice of initial condition enables comparison of the stochastic,  deterministic  and finite difference models with an exact solution on the finite domain. This exact solution to the diffusion equation with diffusivity $D_0$ over the interval $[-L, L]$ is~\cite{Crank1975}
\begin{align}\label{exact-finite}
  &n(x,t) = N_\textrm{tot}\frac{\pi}{4L}\cos\left(\dfrac{\pi x}{2L}\right) \exp\left( - \left(\dfrac{\pi}{2L}\right)^2 D_0 t\right).
\end{align}
\begin{figure}[t]
	\centering   
	\includegraphics[scale=0.324]{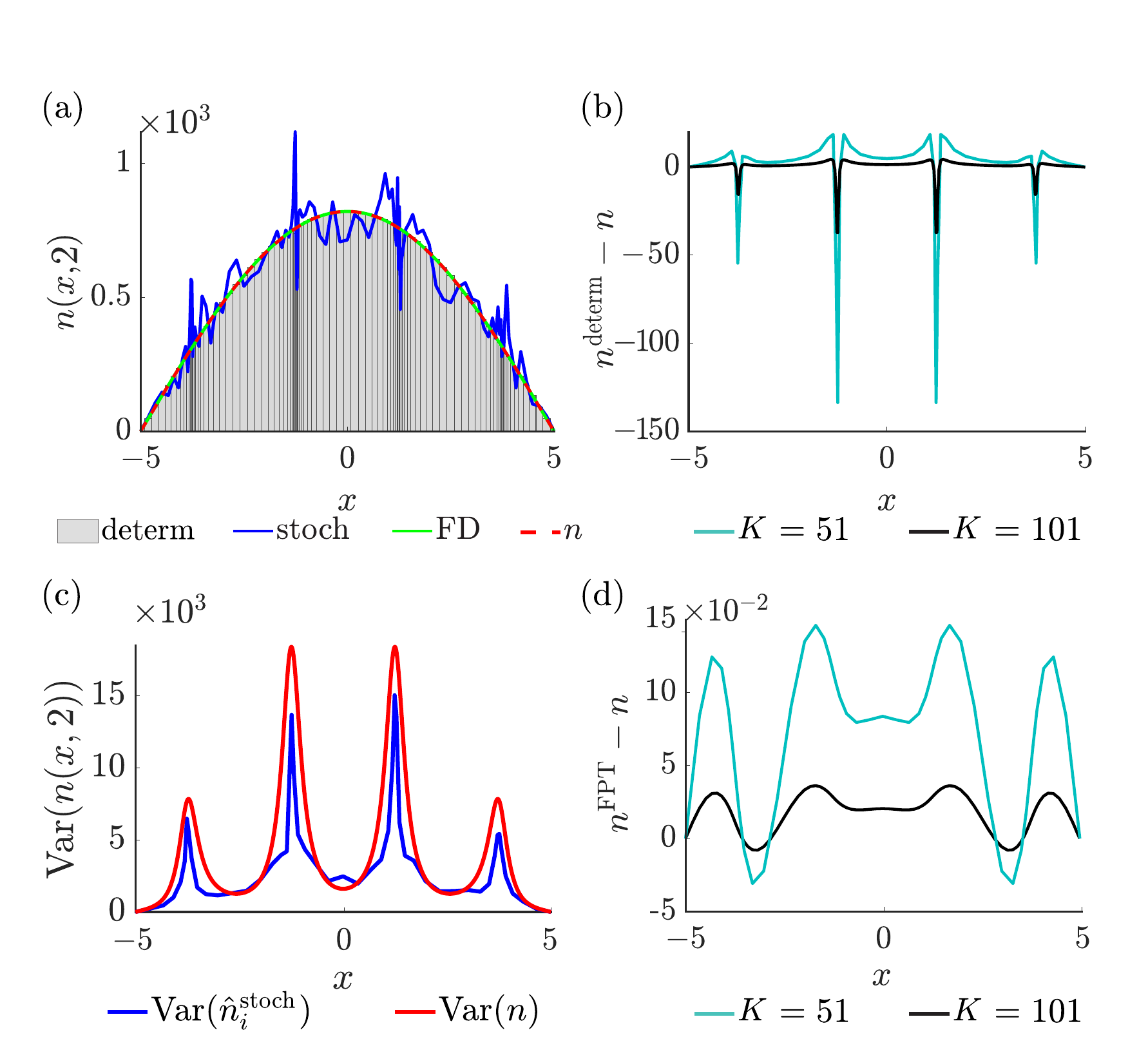}
  \caption{Network with periodic node density, $x(\xi) = \xi + A\sin(k\xi)$  and space-dependent jump probabilities~\eqref{q-const-diff}  (a)~Concentration of the deterministic (gray), stochastic (blue), finite difference (green), and exact (Eq.~\eqref{exact-finite}, red) solutions at $t = 2$  for $K=101$ nodes . (b) The difference between the  solution of the deterministic model  using space-dependent jump probabilities (Eqs~\eqref{space-dep}) and the exact solution at $t = 2$ for $K = 51$ nodes (teal) and $K = 101$ nodes (black). (c) The variance of the stochastic model computed using $100$ stochastic realisations (blue) and as estimated using the exact solution (red) at $t = 2$. (d) The difference between the FPT solution (Eqs~\eqref{yates}) and the exact solution at $t = 2$ for $K = 51$ nodes (teal) and $K = 101$ nodes (black). Parameters: $D_0 = 1$,  $L =5$,  $N_\text{tot}=6000$, $K = 51$ or $K=101$, $k = \frac{8\pi}{2L}$, $A = \frac{4}{5k}$, $\Deltaup t = 1\times 10^{-4}$, $\Deltaup \xi = \frac{2L}{K - 1} = 0.2$  or $0.1$, $M = 150$, $\updelta t = 2 \times 10^{-6}$.}
	\label{fig-periodic-const-diff}
\end{figure}

Figure~\ref{fig-periodic-const-diff}a presents a time snapshot of the  deterministic, stochastic, finite difference and exact solutions for $n(x,t)$. As expected, the finite difference solution closely matches the exact solution. However, the  deterministic  and stochastic models deviate from the exact solutions where the node density $\rho(x)$ is high. In Fig.~\ref{fig-periodic-const-diff}b we present the absolute error between the exact and  deterministic  solutions for $K = 51$ and $K = 101$ at time $t=2$. With more nodes in the network, spacings between nodes become smaller and the discrepancy between the exact and  deterministic  solutions decreases  from a maximum of 17\% relative error with $K=51$ to a maximum of 5\% relative error with $K=101$, consistently with the reaction--diffusion--advection equation being the continuum limit of the  deterministic  model.

Figure~\ref{fig-periodic-const-diff}c presents the variance $\textrm{Var}\left(n_i^\textrm{stoch}(t)\right)$ at time $t = 2$ calculated using $100$ stochastic realisations of this model, each with $N_\text{tot}=1000$ molecules. This is shown alongside the expected variance $\textrm{Var}\left(n(t)\right)$ that is computed with the right-hand expression of Eq.~\eqref{var-n} using the exact solution $n(x,t)$ from Eq.~\eqref{exact-finite} in place of $n_i^{\mathrm{disc}}(t)$. It is clear from Fig.~\ref{fig-periodic-const-diff}c that the variance increases in areas of the network with high node density. This result  may not be intuitive; it emphasises the importance of considering stochastic models, particularly in biological contexts where molecule numbers are small and fluctuations can be large. Mathematically, this result is due to the uneven distribution of nodes in space and the fact that the variance of number of molecules at a node is quadratic in $P_i(t)$ in Eq.~\eqref{var-N}.  The  deterministic  and continuum solutions cannot capture this node-density-dependent behaviour of molecular fluctuations.

Yates et al.~\cite{yates2012going} showed using first passage time (FPT) problems that in order to achieve constant diffusivity in the continuum limit on an irregular network, left and right jump probabilities should be selected as
\begin{equation}\label{yates}
	\tilde{q}_i^{-1}=\dfrac{2D_0\Deltaup t}{h_i + h_{i+1}}\dfrac{1}{h_i}, \quad \text{ and } \quad \tilde{q}_i^{1}=\dfrac{2D_0\Deltaup t}{h_i + h_{i+1}}\dfrac{1}{h_{i+1}},
\end{equation}
where $h_i = x_i - x_{i-1}$. To show that the first passage time jump probabilities $\widetilde{q}_i^{\,\pm 1}$ in Eqs~\eqref{yates} converge to $q_i^{\pm 1}$ in Eqs~\eqref{space-dep} in the limit $\Deltaup\xi\to 0$, we first note that given $h_i=x_i-x_{i-1}$, we have
\begin{align}
	\frac{h_i+h_{i+1}}{2} = \Deltaup x_i &\sim g(\xi_i)\Deltaup \xi, \label{hi+hip1}
	\\h_i = x_i-x_{i-1} &\sim g\big(\xi_{i-1/2}\big)\Deltaup\xi, \label{hi}
	\\h_{i+1} = x_{i+1}-x_i &\sim g\big(\xi_{i+1/2}\big)\Deltaup\xi, \qquad \Deltaup\xi\to 0.\label{hip1}
\end{align}
Substituting Eqs~\eqref{hi+hip1}--\eqref{hip1} into Eqs~\eqref{yates}, expanding $g(\xi_{i\pm 1/2})^{-1} = g(\xi_i)^{-1}\mp g(\xi_i)^{-2}\pd{g(\xi_i)}{\xi}\frac{\Deltaup\xi}{2} + \Order(\Deltaup\xi^2)$ and noting that \mbox{$\pd{}{x}=\frac{1}{g}\pd{}{\xi}$}, we obtain, in the limit $\Deltaup\xi\to 0$,
\begin{align*}
	\widetilde{q}_i^{\,\pm 1} \sim \frac{D_0\Deltaup t}{g(\xi_i)\Deltaup\xi} \frac{1}{g\big(\xi_{i\pm 1/2}\big)\Deltaup\xi} \sim \frac{D_0\Deltaup t}{g(\xi_i){\Deltaup\xi}^2} \left(\frac{1}{g(\xi_i)} \mp \frac{1}{g^2(\xi_i)}\pd{g(\xi_i)}{\xi}\frac{\Deltaup\xi}{2}\right) = \frac{D_0\Deltaup t}{(g\Deltaup\xi)^2}\Big(1\mp\pd{g}{x}\frac{\Deltaup\xi}{2}\Big),
\end{align*}
which matches the expressions $q_i^\pm$ in Eqs~\eqref{space-dep}.
  
We have computed the FPT solution, $n^{\mathrm{FPT}}$, using the jump probabilities from Yates et al.~\cite{yates2012going} corresponding to our periodic network. Figure~\ref{fig-periodic-const-diff}d presents the disparity between the exact solution of Eq.~\eqref{exact-finite} and the FPT solution for $K = 51$ and $K = 101$ at time $t = 2$. As for our  deterministic  model, the error decreases with more nodes in the network. The error in our  deterministic  model is significantly larger than the error using the jump probabilities of Yates et al.~\cite{yates2012going}. However, we emphasise that our approach provides explicit expressions for $q_i^{-1}$, $q_i^0$ and $q_i^1$ in terms of an arbitrary desired advective velocity $v(x)$ and diffusivity $D(x)$ on an arbitrary network (Eqs~\eqref{qi0}--\eqref{qip1}).

\subsection{Signal propagation mechanisms in the osteocyte network}\label{sec:osteocytes}
\begin{figure}
	\centering\includegraphics[scale=0.7]{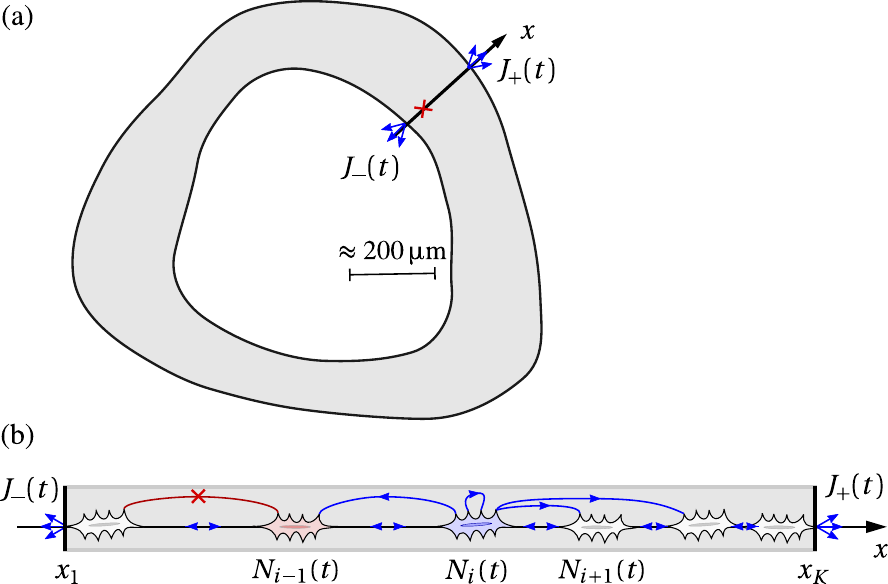}
	\caption{Schematic osteocyte network across the cortical wall of a mouse tibia cross-section. (a) Cortical wall of a mouse tibia cross section (grey) showing the $x$-axis, signalling molecule fluxes $J_-(t)$ and $J_+(t)$ at the endosteal and periosteal bone surfaces, respectively, and the location of osteocyte network damage (red cross). (b) Osteocytes along $x$ connected into a one-dimensional network. The osteocyte in blue represents a stimulated osteocyte generating signalling molecules transfering to neighbouring osteocytes by the jump process. The osteocyte and connection in red represent localised network damage.}\label{fig-1D-Ot-network}
\end{figure}

Signals generated by osteocytes within bone tissue propagate through the osteocyte network to the bone surface, where they induce bone formation or bone resorption~\cite{bonewald-2011,manolagas-parfitt-2013,jilka-noble-weinstein-2013}. In this section we propose a simple, illustrative model of signal generation and propagation through a one-dimensional osteocyte network (Fig.~\ref{fig-1D-Ot-network}). We investigate how signals arriving at the boundaries of the network  are influenced by local perturbations occurring at internal nodes (osteocytes) of the network. The one-dimensional network that we consider may represent osteocytes distributed across the cortical wall of a long bone such as a mouse tibia, between the bone surface of the marrow cavity (endosteal surface) and the outer bone surface (periosteal surface). In the mouse tibia, osteocyte density is greater near the endosteal and periosteal bone surfaces and lowest in the middle of the cortical wall, which is approximately 200\,\um\ thick~\cite{vanTol-etal-2020a,vanTol-etal-2020b}.

We choose the $x$-axis such that the origin represents the middle of the cortical wall, and consider
\begin{align}\label{ot-network-metric}
	x(\xi) =  L\, \text{sign}(\xi) \log(1+\abs{\xi}), \quad\text{so that}\quad \xi(x) = \text{sign}(x)\big(\e^{\abs{x}/L}-1\big), \quad\text{and}\quad g(x) =  L\,\e^{-\abs{x}/L},
\end{align}
 where $L=50\,\um$ is a spatial scale parameter.
With this choice the density of osteocytes increases away from the origin as \mbox{$\rho(x)=\e^{\abs{x}/L}/(L\,\Deltaup\xi)$}. This choice is not intended to be an accurate representation of osteocyte density measurements in mouse tibia~\cite{vanTol-etal-2020b}, but it allows us to qualitatively capture density heterogeneities in real bone while allowing closed-form expressions for $g(x)$ and therefore $D(x)$ and $v(x)$, as well as the long-time behaviour of the solution in infinite space.

Osteocytes are highly connected to neighbouring osteocytes~\cite{kerschnitzki-etal-2013}. We set $q_i^0=0.36, q_i^{\pm 1}=0.16, q_i^{\pm 2}=0.16$ for all internal nodes, with all remaining $q_i^s=0$. This means that each osteocyte is connected to its nearest and next-nearest neighbours, and there is no jump-probability bias.

\begin{figure}[t]
	\centering
	\includegraphics[width=\textwidth]{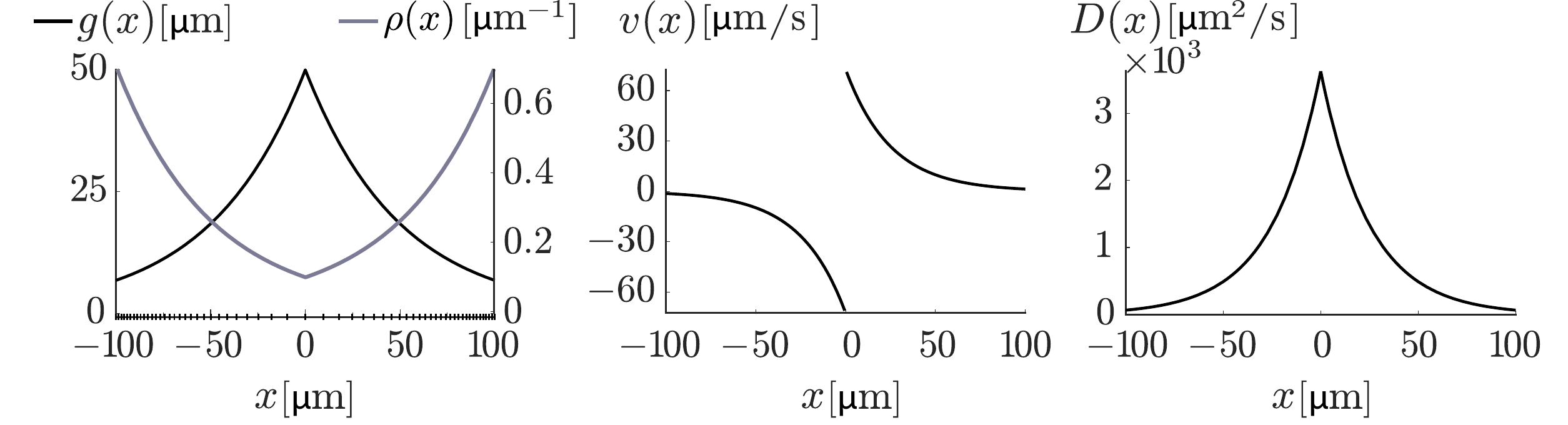}	
	\caption{Network with exponentially increasing node density, $x(\xi) =  L\,\text{sign}(\xi)\log\big(1+\abs{\xi}\big)$: Metric, density, advection, and diffusion functions.  Network node positions are marked on the $x$-axis.  (Parameters: $q_{i}^{\pm 1} = q_{i}^{\pm 2} = 0.16$, $q_{i}^{0} = 0.36$, $K = 61$,  $x_1 = -100\,\um$, $x_K=100\,\um$, $\Deltaup t = 0.025\,\s$.)}
	\label{fig-osteocytes-g-v-D}
\end{figure}

The metric, node density, advection and diffusion functions for this network are presented in Fig.~\ref{fig-osteocytes-g-v-D}.  The jump probabilities and discretisation parameters of the stochastic and deterministic models are chosen so that advective velocities $v(x)$ are of the same order of magnitude as those arising in the fluid flow model of van Tol~et al. (about 20\,$\um/\text{s}$)~\cite{vanTol-etal-2020a}, and diffusive root mean square displacements $\sqrt{2D(x)t\,}$ over one second are of the order of 20--80\,\um. While these values are commensurate with effective fluid flow transport properties~\cite{vanTol-etal-2020a}, other parameter choices could be made to represent other types of signalling through the osteocyte network such as calcium signals~\cite{adachi-etal-2009,lewis-etal-2017}. We note here that the results we present in this section illustrate qualitative behaviour of the model that do not depend on the space and time scales chosen. By rescaling time, space, and densities in Eq.~\eqref{evo-n}, it is possible to modify the scales of the axes in all the figures presented below~\cite{barenblatt}.

The network's metric $g(x)$ and diffusivity $D(x)$ have a cusp at $x=0$ and the advective velocity $v(x)$ is discontinuous at $x=0$.  Because of these singularities in $D(x)$ and $v(x)$, the finite difference solutions of the reaction--diffusion--advection equation are inaccurate compared to those provided by the deterministic model (see Section~\ref{sec:space-indep-jumps}). We therefore omit numerical results from the finite difference simulations in this section.

\begin{figure}[t]
	\centering	\includegraphics[width=\textwidth]{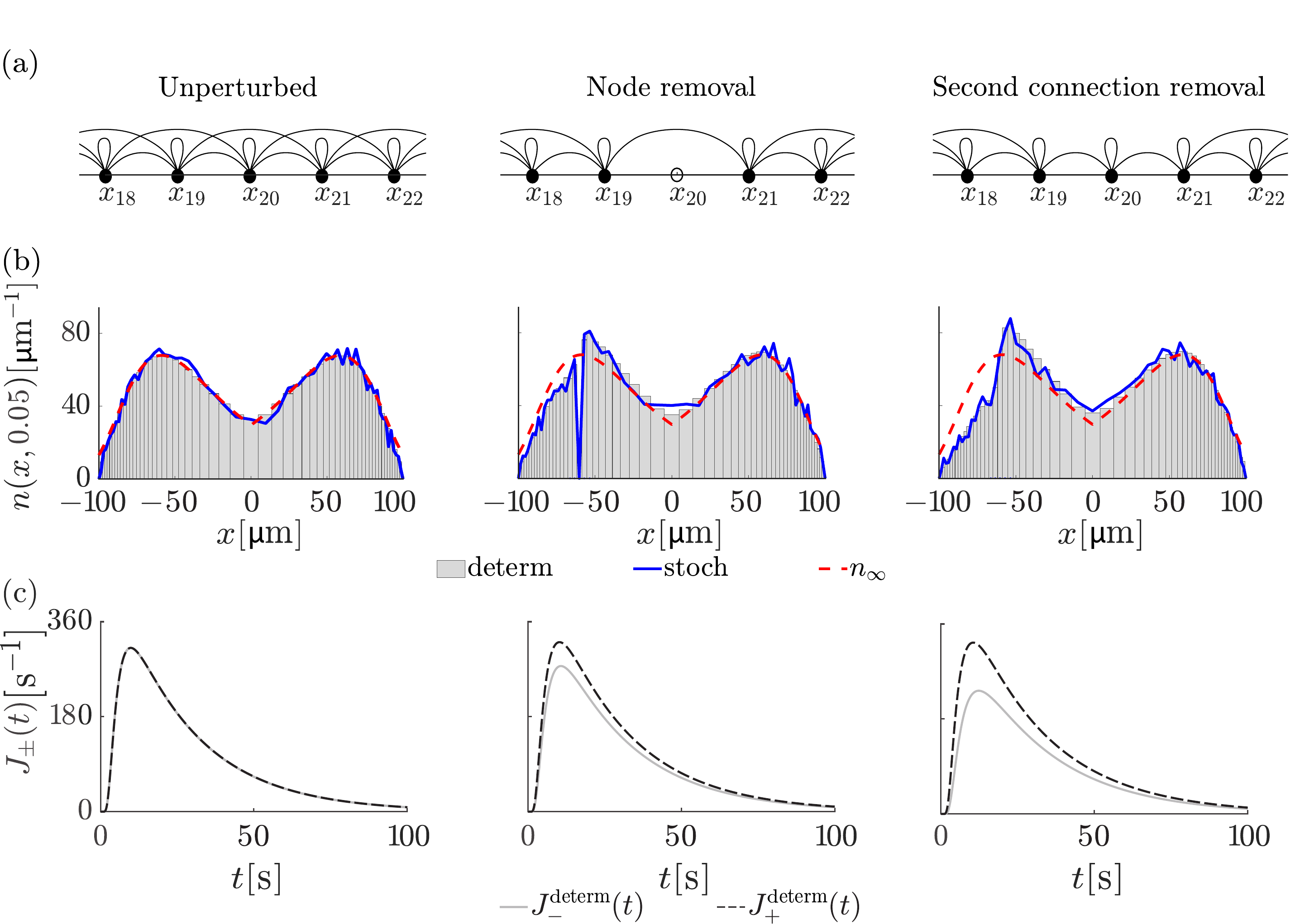}
  \caption{Perturbations of a network with exponentially increasing node density: (a) Schematic of the network depicting the unperturbed case, removal of node $i=20$, and removal of second-neighbour connections at $i=20$. (b) Concentration of  molecules corresponding to the  deterministic  (gray) and stochastic (blue) solutions at time  $t = 2.5\,\s$  for the three networks from an initial impulse with $N_\mathrm{tot}=10,000$ molecules at $x=0$. The presented infinite-space exact solution (red) in all three columns relates to the unperturbed network. (c) Flux corresponding to the  deterministic  solution at left boundary (gray line) and at right boundary (black dashed line) for the three networks. (Parameters as given in Fig.~\ref{fig-osteocytes-g-v-D}).}\label{fig-osteocyte-perturbations}
	\label{f0019}
\end{figure}

We first analyse the robustness of signal propagation under perturbations of the network. We assume no chemical reactions ($F=0$) and consider that an impulse signal is generated at time $t=0$ at the centre of the network ($x=0$). Figure~\ref{fig-osteocyte-perturbations} shows schematics of the considered network perturbations (Fig.~\ref{fig-osteocyte-perturbations}a) and corresponding snapshots of the molecule concentration at time  $t=2.5$  (Fig.~\ref{fig-osteocyte-perturbations}b). Figure~\ref{fig-osteocyte-perturbations}c shows the time signature of the fluxes $J_-(t)$ and $J_+(t)$ of molecules reaching the left and right boundary nodes in response to the initial impulse. These responses represent signals that could induce bone formation or bone resorption at the bone surface.
The left column of Fig.~\ref{fig-osteocyte-perturbations} relates to the unperturbed network. The initial impulse splits into two peaks propagating left and right toward the network boundaries (Fig.~\ref{fig-osteocyte-perturbations}b). While diffusion tends to disperse molecule concentration peaks (cf.~Fig.~\ref{fig-regular}), here the increasing node density encountered by the molecules as they propagate away from the origin counters their dispersion, and creates well-defined outward-moving peaks evolving into a travelling wave of constant shape and height on an infinite domain (see Appendix~\ref{appx:travelling-wave}). 

The centre column of Fig.~\ref{fig-osteocyte-perturbations} relates to a node $i=20$ being removed in the left half of the network. Signalling molecules are still transmitted through to the left boundary $i=1$ due the presence of next-nearest neighbour connections between nodes $i=21$ and $i=19$. However, molecules tend to pile up at node $i=21$ because they have fewer opportunities to leave this node toward the left, resulting in an increased concentration right of the node $i=21$ and a decreased concentration left of the node $i=19$. This results in a lower peak in the flux signature $J_-(t)$ at the left boundary (Fig.~\ref{fig-osteocyte-perturbations}c). The right column of Fig.~\ref{fig-osteocyte-perturbations} relates to second-neighbour connections of node $i=20$ being severed. In this case  molecules progressing toward the left boundary pile up before node $i=20$ since they transfer over the severed region less easily, resulting in an even lower flux $J_-(t)$ at the left boundary. These local changes in the network may represent micro-damage in bone tissue such as osteocyte death or micro-cracks severing connections. Micro-damage in bone is removed by a bone remodelling process initiated at the bone surface~\cite{martin-2003,manolagas-parfitt-2013}. Differences seen in our simulations between the signal received at the left and at the right boundaries due to an off-centre damaged osteocyte could thus enable localisation of the damage and help guide repair processes, such as triggering a bone remodelling event to be initiated closest to the damage.

\begin{figure}[t]
	\centering
	\includegraphics[scale=0.326]{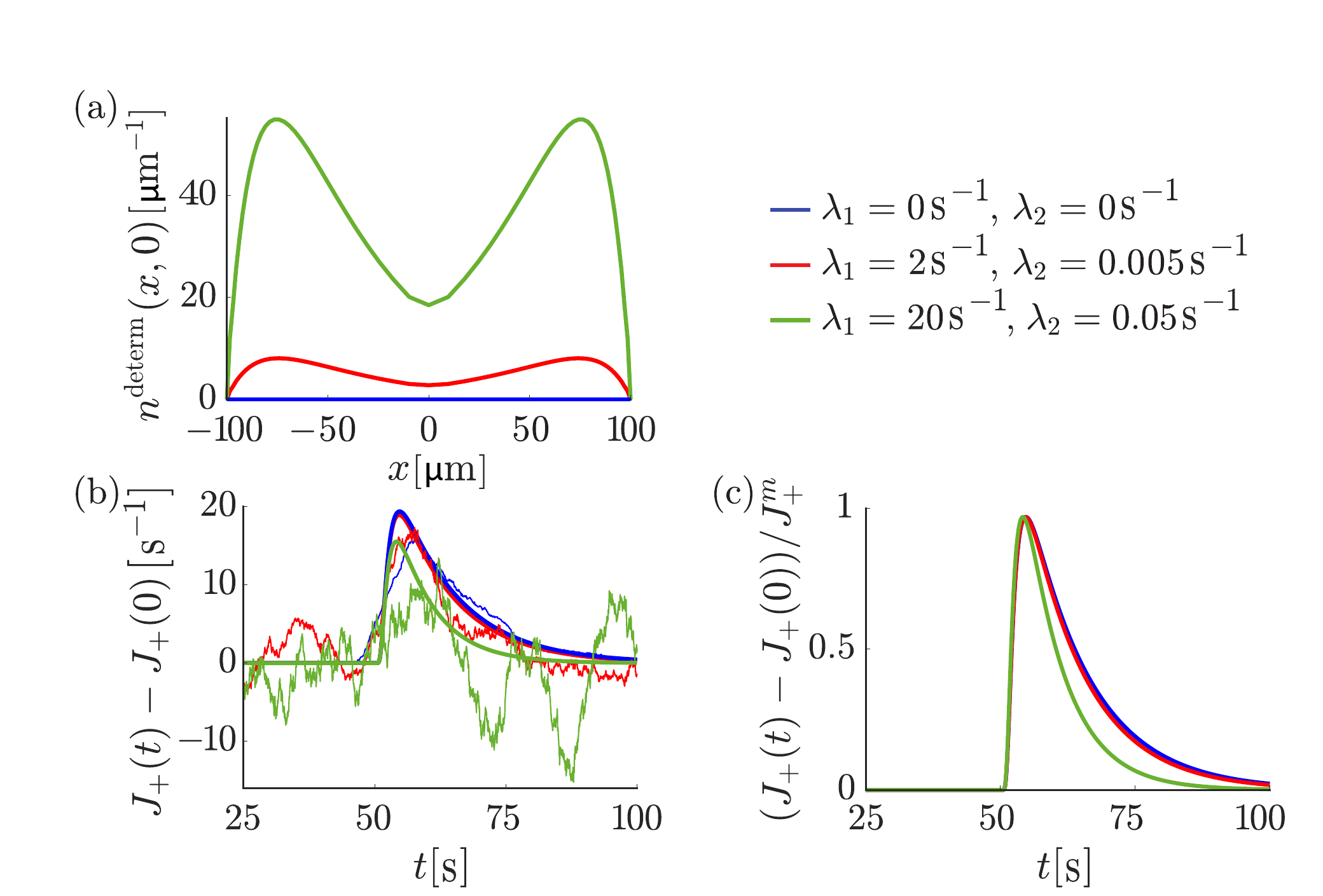}	
  \caption{Influence of reactions on signalling within a network with exponentially increasing node density for the three cases of no reactions ($\lambda_1 = 0$ and $\lambda_2 = 0$, blue curves), reaction rates of  $\lambda_1 = 2\,\s^{-1}$ and $\lambda_2 = 0.005\,\s^{-1}$  (red curves), and reaction rates of  $\lambda_1 = 20\,\s^{-1}$ and $\lambda_2 = 0.05\,\s^{-1}$  (green curves): (a) Concentration of molecules at steady state in the  deterministic  model; (b)  Flux of the deterministic and stochastic models  at the right boundary starting from the steady state and with an impulse of $600$ molecules added at $x=0$ at time  $t=50\,\s$. The curves are shifted by $J_+(0)$ so they start from the same baseline. (c)  Flux of the deterministic model  at the right boundary as in Fig.~\ref{fig-osteocytes-ss-impulse}b, where the curves are shifted by $J_+(0)$ and scaled by $J_{+}^{m} = {\displaystyle \max_t}{J_{+}(t) - J_+(0)}$. (Parameters as given in Fig.~\ref{fig-osteocytes-g-v-D}. Stochastic fluxes are averaged over a time window of $10\,\s$).}
	\label{fig-osteocytes-ss-impulse}
\end{figure}

We now investigate the influence of localised changes in molecule generation by the osteocyte network, and how these changes affect the signals received at the bone surface (network boundaries). We assume a steady, homeostatic state, in which each osteocyte generates a background of signalling molecules with rate $\lambda_1$ (number per unit time). The molecules are also assumed to degrade with a first-order reaction rate $\lambda_2$ (probability per unit time), so that
\begin{align}
	C(N) = \lambda_1, \quad E(N)=-\lambda_2 N, \quad F(N) = \lambda_1-\lambda_2 N.
\end{align}
This homeostatic state models osteocytes subjected to normal mechanical stimulation. The constant creation and elimination of molecules as they diffuse through the network generates a fluctuating quasi-steady state in the stochastic model, with an average concentration of molecules along $x$ that is affected by the node density and the absorbing boundary conditions, as well as the strength of reaction rates $\lambda_1$ and $\lambda_2$ (Fig.~\ref{fig-osteocytes-ss-impulse}a). In the stochastic model, the fluxes $J_\pm(t)$ of signalling molecules at the network boundaries are highly fluctuating quantities in the presence of stochastic reactions because we assume only one reaction per node per time step $\Deltaup t$. We therefore report the flux moving average values over a  time window of 10\,s.

If an osteocyte at the centre of the network produces an additional impulse in the number of molecules at a certain time, this impulse will split into two peaks propagating left and right toward the boundaries and generate a peak in $J_\pm(t)$ similarly to Fig.~\ref{fig-osteocyte-perturbations}. However, the intensity and duration of this impulse response depends on the reaction rates $\lambda_1$ and $\lambda_2$ (Fig.~\ref{fig-osteocytes-ss-impulse}). Indeed, the continual creation and degradation of molecules at each node acts as a chemical reservoir that tends to equilibrate the number of molecules at each node~\cite{van1992stochastic,lauffenburger-linderman-1993}. Higher reaction rates lead to a faster return to equilibrium. This results in a more attenuated response (Fig.~\ref{fig-osteocytes-ss-impulse}b), but also in less broadening of the signal in time (Fig.~\ref{fig-osteocytes-ss-impulse}c).

\begin{figure}[t]
	\centering
		\hspace{2cm}\includegraphics[scale=0.40]{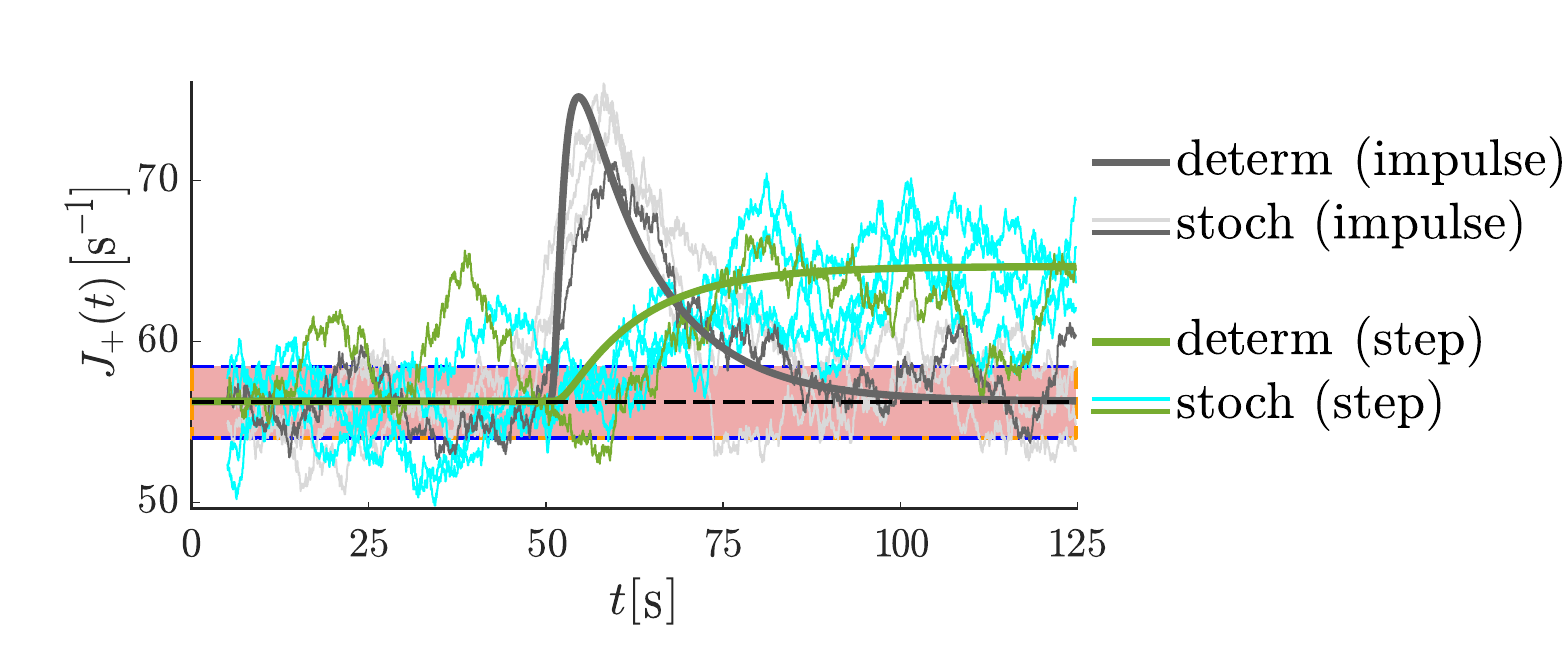}
  \caption{Response to impulse and step signals from steady state within a network with exponentially increasing node density:  flux of the deterministic and stochastic models  at the right boundary for an impulse of $600$ molecules added at $x=0$ at  $t=50\,\s$  (dark gray and light gray), and a step increase in $\lambda_1$ from  $2\,\s^{-1}$ to $20\,\s^{-1}$  at $x = 0$ for  $t \geq 50\,\s$  (dark green and cyan). Each case includes two stochastic realisations. The dashed black line indicates the mean of the stochastic flux at the steady state, and the blue dashed lines indicate one standard deviation from the mean. (Parameters as given in Fig.~\ref{fig-osteocytes-g-v-D} with  $\lambda_1 = 2\,\s^{-1}$ and $\lambda_2 =0.005\,\s^{-1}$  for all nodes except at $x=0$, and a time averaging window for the stochastic flux of  $10\,\s$.)}\label{fig-osteocyte-signals}
  
\end{figure}

Figure~\ref{fig-osteocytes-ss-impulse}b shows that the intensity of the impulse response $J_+(t)$ at the network boundary may be of similar order of magnitude as inherent fluctuations of $J_+(t)$, and may therefore not be easily detected. In Fig.~\ref{fig-osteocyte-signals} we show several stochastic realisations of the signal response $J_+(t)$ received at the right boundary of the network following either an impulse increase in molecules at the central node at time  $t=50\,\s$, or a step increase in the rate of molecule production $\lambda_1$ at the central node for  $t\geq 50\,\s$, starting from a (fluctuating) initial steady state at $t=0$. The first signal production represents an osteocyte generating a sudden, short-lived burst of signalling molecules, such as due to apoptosis~\cite{jilka-noble-weinstein-2013}. The second signal production respresents a continued response in time, such as due to the detection of local micro-damage, or mechanical overstimulation. In the  deterministic  model, both types of signals result in a noticeable (average) response, but in the stochastic model, some stochastic fluctuations in steady state may be misinterpreted as a response to an impulse signal. In contrast, the response to the step increase in production rate generates a stronger signal-to-noise ratio in the stochastic model.

\section{Conclusion} 
In this paper we have presented  stochastic and deterministic versions of  random walk models for the reaction and propagation of signalling molecules on one-dimensional spatial networks with arbitrary node placement and connectivity. We have derived a reaction--diffusion--advection differential equation that is the continuum limit for these models. Solutions of the reaction--diffusion--advection equation capture the average behaviour of the stochastic model well when spacings  and connections  between nodes are small, or equivalently, when the network is dense.

The continuum limit elucidates how signalling molecules propagating through spatial networks are transported in physical space through diffusion and advection properties that depend on local node density and connectivity within the network. The relationships we derived between spatial network properties and diffusion and advection can help us understand average transport properties in  physical communication networks. Our results highlight that the concentration of molecules in space and their diffusivity depend strongly on the node density of the network. Node density can thus have important consequences for the development of spatial patterns of morphogens in addition to the influence of network connectivity~\cite{nakao-mikhailov-2010}. Node density impacts the efficiency with which molecular factors diffuse and establish in biological tissues to perform their function.  As already mentioned in the introduction,  in plants, for example, the functional behaviour of the network of cell-to-cell communication controls average tissue-scale behaviours such as tissue growth and defense mechanisms~\cite{maule-etal-2012}, as well as patterning and organ development~\cite{jackson-etal-bassel-2019}. In distributed networks such as cardiovascular, respiratory and root systems, spatial network properties will influence average propagation times of nutrients and waste products~\cite{banavar-maritan-rinaldo-1999}.  In the example of bone tissue, the osteocyte network regulates the mineralisation of bone, which provides compressive strength and fast calcium storage and retrieval~\cite{atkins-findlay-2012,heveran-boerckel-2023}. Experimental data on osteocyte networks and mineral density exhibit patterns suggesting complex diffusion and reaction behaviours of mineral precursors and inhibitors through the osteocyte network~\cite{ayoubi-etal-2021}, that could lead to future analyses based on our modelling approach.

While the  deterministic discrete  model and continuum reaction--diffusion--advection equation provide accurate representations of the average behaviour of the stochastic random walk model, they do not account for fluctuations in molecule concentrations inherent to stochasticity in both transport and chemical reactions. The intensity of fluctuations was found to depend strongly on the network's node density. This could affect signal-to-noise ratios significantly in signals propagating through cellular networks, in which a baseline of stochasticity can be large. Some chemical reaction behaviours such as regulatory switches and Turing patterns can be triggered by fluctuations~\cite{gardner-cantor-collins-2000,cao-erban-2014}. The consideration of the spatial arrangement of nodes of the network in stochastic signal propagation in these situations could lead to new behaviours that are of high interest.

Numerically solving inhomogeneous reaction--diffusion--advection equations with rapidly varying diffusivity and advective velocity can be challenging. Sophisticated conservative finite-volume methods  or finite element methods  are often used for this purpose, since standard finite difference methods can fail to provide accurate solutions, as illustrated by our results when the metric is close to zero and the diffusivity and advective velocity have singularities. However, our results show that the  deterministic  discrete model can be used as an effective, conservative, and explicit  numerical method for solving the reaction--diffusion--advection equation in all situations. Recent work has utilised a similar approach to solve nonlinear advection--diffusion equations on regular grids~\cite{angstmann-henry-etal-2019}. Our approach enables us to obtain expressions for nearest-neighbour jump probabilities that will achieve desired inhomogenous advective velocity and diffusivity in the continuum equation for a given, arbitrary grid.

 We illustrated a particular application of our model to the osteocyte network in bone. In this system, positional information about where signals originate from deep inside the network in response to damage is important so that this damage can be repaired~\cite{martin-2003}.  Signals generated by osteocytes within bone tissue in response to damage propagate to the bone surface, where they induce  repair through bone resorption and bone formation. The spatial location of these signals is critical, so that bone resorption is not initiated where bone tissue is needed most, and so that micro-damage within bone tissue induces targeted remodelling. Our investigation of networks in which internal nodes are perturbed shows that signals received at the boundaries of the network can indicate the location of the perturbation. However, these signals may also be difficult to detect depending on the nature of signal generation and the presence of chemical reactions. The considered reactions contribute significantly to noise in the signal received at the network boundaries, with noise increasing with reaction rate. The noise can result in an impulse of signalling molecules at the centre of the network being difficult to detect at the network boundaries. However, a step signal, for example an increase in the reaction rate at the central node of the network, generates a stronger signal-to-noise ratio at the network boundaries.  Our model of the osteocyte network is currently limited by only accounting for one spatial dimension. In a one-dimensional network, localised damage is always  necessarily  traversed by propagating signals. In higher dimensions, localised damage may affect propagating signals less as there are more paths available to bypass the damage. Experimental datasets of 3D osteocyte network architectures from confocal microscopy have been employed to simulate fluid flow through the lacuno-canalicular network in which osteocytes reside~\cite{vanTol-etal-2020a,vanTol-etal-2020b}. While our model considers a simplified, one-dimensional situation, it still represents important mechanisms by which the osteocyte network is believed to operate in real bone tissues. Some of these mechanisms have not been previously considered, such as the influence of disruptions to the network, chemical reactions, and stochasticity.

 The extension of our model to higher dimensions and to dynamic, adaptive networks~\cite{porter-gleeson-2016} is of strong interest as it would enable more comprehensive modelling of several network systems. Deriving continuum limits for arbitrary networks in higher dimensions is a nontrivial extension but it can be anticipated that average transport properties of signals may also be governed by reaction--diffusion--advection equations, since these types of equations express conservation laws that hold at the discrete level~\cite{evans-morriss,vandenHeuvel-etal-2024}.  In bone, such models may be able to further our understanding of how signals propagating through the osteocyte network contribute to bone formation and resorption processes, and how osteocytes set the mechanical memory of bone~\cite{lerebours-buenzli-2016}.

Further avenues for future work include the consideration of more detailed transport problems, such as propagation of signalling molecules along the connections between the network's nodes, which result in partial differential equations on networks~\cite{porter-gleeson-2016}.

\subsubsection*{Data accessibility} Key algorithms used to generate results are freely available on Github at \url{https://github.com/prbuen/Mehrpooya2023_RandomWalk1DNetwork}.

\subsubsection*{Acknowledgments}PRB wishes to thank Richard Weinkamer for helpful discussions. PRB acknowledges support from the Australian Research Council (DP190102545). AM, VJC and PRB acknowledge support from the Max Planck Queensland Centre for the Materials Science of Extracellular Matrices.

\appendix
\section{Long-time limit of peak propagation in the osteocyte network}\label{appx:travelling-wave}
\renewcommand{\theequation}{\thesection\ \arabic{equation}}
 In this appendix we seek a long-time solution of signal propagation through the one-dimensional network given by Eqs~\eqref{ot-network-metric}, in which node density increases symmetrically away from the origin as $\rho(x)=\e^{\abs{x}/L}/(L\Delta\xi)$, where $L$ is a spatial scale parameter taken to be 50\,\um\ in Section~\ref{sec:osteocytes}. Numerical simulations of the propagation of an initial impulse at the origin  show that the impulse splits into two peaks propagating toward the left and the right. Here, we characterise these peaks in the long-time limit in infinite space as travelling waves propagating without change in shape, but with a decreasing speed.

\begin{figure}
	\centering\includegraphics[scale=0.85]{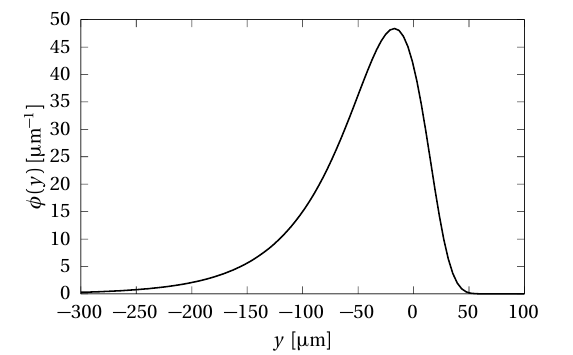}
	\caption{Long-time wave profile $\phi(y)$ given by Eq.~\eqref{wave-profile} with $N_\text{tot}=10,000$ and $L=50\,\um$. The wave is generated after an impulse initial condition at the origin, and it travels toward the right with speed $v(t)=  L/(2t)$. Nodes of the network are distributed according to Eqs~\eqref{ot-network-metric}.}\label{fig-wave-peak}
\end{figure}

Since the solution is symmetrical with respect to the origin, we focus on the right-propagating wave only ($x>0$). In the long-time limit $t\to\infty$, the solution is nonzero only in the vicinity of the peak, for $x\gg 1$.  The solution to the diffusion--advection equation~\eqref{evo-n} for molecule concentration $n_\infty(x,t)$ in infinite space in the absence of reactions ($F(N)=0$) and with unbiased, constant jump probabilities ($\widetilde{v}=0$) is given by the metric-scaled Gaussian in Eq.~\eqref{n-infty}. For the choice of network considered here, the metric is $g(x)=L\e^{-\abs{x}/L}$, so that:
\begin{align}
	n_\infty(x,t) &= \frac{N_\text{tot}\ \e^{x/L}}{ L\sqrt{4\pi \widetilde{D} t}} \exp\left\{ - \frac{(\e^{x/L}-1)^2}{4\widetilde{D} t}\right\}
	\\&\sim \frac{N_\text{tot}}{ L\sqrt{\pi}} \frac{\e^{x/L}}{\sqrt{4\widetilde{D} t}}\exp\left\{-\left(\frac{\e^{x/L}}{\sqrt{4\widetilde{D}t}}\right)^2\right\}, \qquad x\gg 1,
\end{align}
 where $\widetilde{D}$ (in units of $\s^{-1}$) is the diffusivity along the dimensionless $\xi$-axis and is related to the discrete models by $\widetilde{D}=\langle s^2\rangle\Deltaup\xi^2/(2\Deltaup t)$ which involves the second moment $\langle s^2\rangle$ of the jump probabilities, see Eq.~\eqref{Dtilde-vtilde}.  
Rewriting $\e^{x/L}/\sqrt{4\widetilde{D}t} = \exp\left(  x/L - \log(\sqrt{4\widetilde{D}t})\right)$, we see that the asymptotic behaviour of $n_\infty(x,t)$ can thus be expressed as a travelling wave
\begin{align}\label{wave-profile}
	n_\infty(x,t) \sim \phi\left( x -  L\log(\sqrt{4\widetilde{D}t})\right),\quad x\gg 1,\quad \text{with}\quad \phi(y) = \psi\big(\e^{y/L}\big), \qquad \psi(z) = \frac{N_\text{tot}}{ L\sqrt{\pi}} z\e^{-z^2}.
\end{align}
The wave profile $\phi(y)$ is shown in Figure~\ref{fig-wave-peak}.  The peak location of the profile along $y$ is such that $\psi'(z)=0$ where $z=\e^{y/L}$, which gives $z=1/\sqrt{2}$ so $y= L \log(z)=- L\log(\sqrt{2})$. The moving peak location $x_\text{peak}(t)$ along the $x$ axis is such that $y=x_\text{peak}(t)- L\log(\sqrt{4\widetilde{D}t})=-L\log(\sqrt{2})$, which gives 
\begin{align}
	x_\text{peak}(t) =  L \log(\sqrt{4\widetilde{D}t})- L\log(\sqrt{2}) =  L\log(\sqrt{2\widetilde{D}t}).
\end{align}
The wave profile thus travels at a decreasing speed  $v(t) =\displaystyle \td{x_\text{peak}(t)}{t} =  \frac{L}{2t}$. The peak height is given by
\begin{align}
	\psi(1/\sqrt{2}) =\frac{N_\text{tot}}{ L\sqrt{2\pi}}\,\e^{-1/2}.
\end{align}
Thus, the exponential increase of node density with $x$ in this network counters precisely the diffusion-driven dispersion of the initial impulse, and gives rise to a travelling wave of constant profile, but decreasing speed.

\end{document}